\newcommand{\zavalasource}{CEERS-DSFG-1}
\newcommand{\donnansource}{CEERS-93316}
\newcommand{\finkelsteinsource}{Maisie's galaxy}
\newcommand{\um}{$\mu$m}

\documentclass[twocolumn]{aastex631}

\usepackage[T1]{fontenc}
\usepackage[utf8]{inputenc}

\shorttitle{\rm DSFGs masquerading as ultra-high redshift galaxies}
\shortauthors{The CEERS collaboration}
\graphicspath{{./}{figures/}}

\begin{document}



\title{Dusty Starbursts Masquerading as Ultra-high Redshift Galaxies in JWST CEERS Observations}

\suppressAffiliations

\author[0000-0002-7051-1100]{Jorge A. Zavala}
\affiliation{National Astronomical Observatory of Japan, 2-21-1 Osawa, Mitaka, Tokyo 181-8588, Japan}
\author[0000-0003-3441-903X]{V\'eronique Buat}
\affiliation{Aix Marseille Univ, CNRS, CNES, LAM Marseille, France}
\author[0000-0002-0930-6466]{Caitlin M. Casey}
\affiliation{Department of Astronomy, The University of Texas at Austin, Austin, TX, USA}
\author[0000-0001-8519-1130]{Steven L. Finkelstein}
\affiliation{Department of Astronomy, The University of Texas at Austin, Austin, TX, USA}
\author[0000-0002-4193-2539]{Denis Burgarella}
\affiliation{Aix Marseille Univ, CNRS, CNES, LAM Marseille, France}
\author[0000-0002-9921-9218]{Micaela B. Bagley}
\affiliation{Department of Astronomy, The University of Texas at Austin, Austin, TX, USA}
\author[0000-0003-0541-2891]{Laure Ciesla}
\affiliation{Aix Marseille Univ, CNRS, CNES, LAM Marseille, France}
\author[0000-0002-3331-9590]{Emanuele Daddi}
\affiliation{Universit\'e Paris-Saclay, Universit\'e Paris Cit\'e, CEA, CNRS, AIM, 91191, Gif-sur-Yvette, France}
\author[0000-0001-5414-5131]{Mark Dickinson}
\affiliation{NSF's National Optical-Infrared Astronomy Research Laboratory, 950 N. Cherry Ave., Tucson, AZ 85719, USA}
\author[0000-0001-7113-2738]{Henry C. Ferguson}
\affiliation{Space Telescope Science Institute, Baltimore, MD, USA}
\author[0000-0002-3560-8599]{Maximilien Franco}
\affiliation{Department of Astronomy, The University of Texas at Austin, Austin, TX, USA}

\author[0000-0002-2640-5917]{E.~F. Jim\'enez-Andrade}
\affiliation{Instituto de Radioastronomía y Astrofísica, UNAM Campus Morelia, Apartado postal 3-72, 58090 Morelia, Michoacán, México}

\author[0000-0001-9187-3605]{Jeyhan S. Kartaltepe}
\affiliation{Laboratory for Multiwavelength Astrophysics, School of Physics and Astronomy, Rochester Institute of Technology, 84 Lomb Memorial Drive, Rochester, NY 14623, USA}
\author[0000-0002-6610-2048]{Anton M. Koekemoer}
\affiliation{Space Telescope Science Institute, 3700 San Martin Drive, Baltimore, MD 21218, USA}

\author[0000-0002-9466-2763]{Aur{\'e}lien Le Bail}
\affil{Universit{\'e} Paris-Saclay, Université Paris Cit{\'e}, CEA, CNRS, AIM, 91191, Gif-sur-Yvette, France}

\author[0000-0001-7089-7325]{E.~J. Murphy}
\affiliation{National Radio Astronomy Observatory, 520 Edgemont Road, Charlottesville, VA 22903, USA}

\author[0000-0001-7503-8482]{Casey Papovich}
\affiliation{Department of Physics and Astronomy, Texas A\&M University, College Station, TX, 77843-4242 USA}
\affiliation{George P.\ and Cynthia Woods Mitchell Institute for Fundamental Physics and Astronomy, Texas A\&M University, College Station, TX, 77843-4242 USA}
\author[0000-0002-8224-4505]{Sandro Tacchella}
\affiliation{Kavli Institute for Cosmology, University of Cambridge, Madingley Road, Cambridge, CB3 0HA, UK}\affiliation{Cavendish Laboratory, University of Cambridge, 19 JJ Thomson Avenue, Cambridge, CB3 0HE, UK}
\author[0000-0003-3903-6935]{Stephen M.~Wilkins} %
\affiliation{Astronomy Centre, University of Sussex, Falmer, Brighton BN1 9QH, UK}
\affiliation{Institute of Space Sciences and Astronomy, University of Malta, Msida MSD 2080, Malta}
\author[0000-0002-6590-3994]{Itziar Aretxaga} %
\affiliation{Instituto Nacional de Astrof\'isica, \'Optica y Electr\'onica, Luis Enrique Erro 1, Tonantzintla CP 72840, Puebla, M\'exico}
\author[0000-0002-2517-6446]{Peter Behroozi}
\affiliation{Department of Astronomy and Steward Observatory, University of Arizona, Tucson, AZ 85721, USA}
\affiliation{Division of Science, National Astronomical Observatory of Japan, 2-21-1 Osawa, Mitaka, Tokyo 181-8588, Japan}
\author[0000-0002-6184-9097]{Jaclyn B. Champagne}
\affiliation{Department of Astronomy, The University of Texas at Austin, Austin, TX, USA}

\collaboration{102}{and The CEERS Team:}
\affiliation{Space Telescope Science Institute, 3700 San Martin Drive, Baltimore, MD 21218, USA}

\author[0000-0003-3820-2823]{Adriano Fontana}
\affiliation{INAF - Osservatorio Astronomico di Roma, via di Frascati 33, 00078 Monte Porzio Catone, Italy}

\author[0000-0002-7831-8751]{Mauro Giavalisco}
\affiliation{University of Massachusetts Amherst, 710 North Pleasant Street, Amherst, MA 01003-9305, USA}

\author[0000-0002-5688-0663]{Andrea Grazian}
\affiliation{INAF--Osservatorio Astronomico di Padova, Vicolo dell'Osservatorio 5, I-35122, Padova, Italy}

\author[0000-0001-9440-8872]{Norman A. Grogin}
\affiliation{Space Telescope Science Institute, 3700 San Martin Drive, Baltimore, MD 21218, USA}

\author[0000-0001-8152-3943]{Lisa J. Kewley}
\affiliation{Harvard-Smithsonian Center for Astrophysics, 60 Garden Street, Cambridge, MA 02138, USA}

\author[0000-0002-8360-3880]{Dale D. Kocevski}
\affiliation{Department of Physics and Astronomy, Colby College, Waterville, ME 04901, USA}

\author[0000-0002-5537-8110]{Allison Kirkpatrick}
\affiliation{Department of Physics and Astronomy, University of Kansas, Lawrence, KS 66045, USA}

\author[0000-0003-3130-5643]{Jennifer M. Lotz}
\affiliation{Gemini Observatory/NSF's National Optical-Infrared Astronomy Research Laboratory, 950 N. Cherry Ave., Tucson, AZ 85719, USA}

\author[0000-0001-8940-6768]{Laura Pentericci}
\affiliation{INAF - Osservatorio Astronomico di Roma, via di Frascati 33, 00078 Monte Porzio Catone, Italy}

\author[0000-0003-4528-5639]{Pablo G. P\'erez-Gonz\'alez}
\affiliation{Centro de Astrobiolog\'{\i}a (CAB/CSIC-INTA), Ctra. de Ajalvir km 4, Torrej\'on de Ardoz, E-28850, Madrid, Spain}

% ...





\begin{abstract}
Lyman-break galaxy (LBG) candidates at $z\gtrsim10$
are rapidly being identified in JWST/NIRCam observations. Due to the (redshifted) break produced by neutral hydrogen absorption of rest-frame UV photons, these sources are expected to drop out in the bluer filters while being well detected in redder filters.
However, here we show that dust-enshrouded star-forming galaxies at lower redshifts ($z\lesssim7$) may also mimic the near-infrared colors of $z>10$ LBGs, representing potential contaminants in LBG candidate samples.
First, we analyze \zavalasource, a NIRCam dropout undetected in the F115W and F150W filters but detected at longer wavelengths.
Combining the JWST data with (sub)millimeter constraints, including deep NOEMA interferometric observations, we show that this source is a dusty star-forming galaxy (DSFG) at $z\approx5.1$.
We also present a tentative $2.6\sigma$ SCUBA-2 detection at 850$\,\mu\rm m$ around a recently identified $z\approx16$ LBG candidate in the same field and show that, if the emission is real and associated with this candidate,
the available photometry is consistent with a $z\sim5$ dusty galaxy with strong nebular emission lines despite its blue near-IR colors. Further observations on this candidate are imperative to mitigate the low confidence of this tentative emission  and its positional uncertainty.
Our analysis shows that robust (sub)millimeter detections of NIRCam dropout galaxies likely imply $z\sim4-6$ redshift solutions,
where the observed near-IR break would be the result of a strong rest-frame optical Balmer break combined with high dust attenuation and strong nebular line emission, rather than the rest-frame UV Lyman break.
This provides evidence that DSFGs may contaminate searches for ultra-high redshift LBG candidates from JWST observations.
\end{abstract}



\section{Introduction} \label{sec:intro}

The superb sensitivity of the JWST coupled with its high angular resolution and its near infrared (near-IR) detectors \citep{Rigby2022a} provides
a unique view of the Universe previously invisible to other telescopes, from nearby star-forming regions to the farthest, faintest galaxies ever found. In the field of extragalactic astronomy, JWST allows us to extend the Lyman-break galaxy (LBG) selection technique beyond $z\gtrsim11$, the redshift at which the Lyman break is redshifted beyond the reach of {\it Hubble Space Telescope} coverage (Hubble serving as the previous workhorse instrument for the identification of such galaxies before the arrival of JWST; see reviews by \citealt{Finkelstein2016,Stark2016,Robertson2021} and references therein).

The identification of very high-redshift LBGs has strong implications for our understanding of galaxy formation and evolution. For example, the confirmation of large numbers of $z>11$ galaxies can provide strong constraints on the formation epoch of the first galaxies and their star formation efficiencies. Their existence can shed light on the dark matter halo mass function in the early Universe, particularly with the presence of very luminous sources found $\lesssim$400\,Myr after the Big Bang (e.g. \citealt{Behroozi2019a}).

In the first few days after the release of JWST observations, an increasing number of samples of LBG candidates at $z\gtrsim10$ were identified (\citealt{Adams2022a,Atek2022,Castellano2022,Donnan2022,Finkelstein2022,Harikane2022,Naidu2022,Yan2022}). The abundance and masses of these sources start to be in tension with the predictions from most galaxy formation models (\citealt{Boylan-Kolchin2022,Finkelstein2022,Lovell2022}).
Nevertheless, the observed colors for some of these very high-redshift candidates may be degenerated with other populations of galaxies at lower redshifts. This results from confusion between the Lyman-$\alpha$ forest break at $z > 12$ with the Balmer and the 4000~\AA\ breaks combined with dust attenuation and/or strong nebular emission.   This means that dusty star-forming galaxies (DSFGs) at significantly lower redshifts ($z\lesssim6-7$) can mimic the JWST/NIRCam colors of $z\gtrsim10$ LBGs, particularly in the shortest-wavelength filters.  While models tend to assume these galaxies are universally red in color, thus distinguishable from the typically very blue LBGs, the complex environments of the ISM within DSFGs plus contamination from nebular emission lines could lead to a mix of observed near-IR colors \citep{howell2010a,casey2014b}, further obfuscating the secure identification of ultra-high redshift LBGs.  The phenomenon of DSFGs contaminating high-redshift LBG searches is, in fact, not new to JWST, as often $z\sim2-3$ DSFGs were found to contaminate $z\sim6-8$ LBG samples selected by {\it HST} (e.g. \citealt{Dunlop2007}); here, both the contaminants (DSFGs at $z\sim4-6$) and LBG targets ($z\sim10-18$) for JWST have shifted to higher redshifts.

The secure identification of LBGs is thus important not only to quantify the contamination fraction in $z\gtrsim10$ LBG samples (which could relax the observed tension between observations and model predictions) but also to constrain the volume density and physical properties of early massive quiescent galaxies and high-redshift DSFGs, an important step towards our ultimate goal of understanding galaxy formation and evolution. However, distinguishing these galaxies from other populations has proven
challenging and requires spectroscopic or multi-wavelength observations probing the older stellar populations (for the quiescent systems) or the dust thermal emission (for DSFGs).

Here, we use JCMT/SCUBA-2 $850\,\rm\mu m$ and NOEMA 1.1\,mm interferometric observations, in combination with the JWST data from the Cosmic Evolution Early Release Science (CEERS) Survey (\citealt{Finkelstein2017a}; \citealt{Bagley2022}; Finkelstein et al., in prep), to search for dust emission around $z>12$ galaxy candidates and NIRCam dropout sources.
We report on a galaxy, \zavalasource, that is undetected in the NIRCam F115W and F150W filters, but whose photometric redshift is well constrained to be around $z=5$ after including (sub)millimeter data.
We also study the $z\sim16.7$ candidate, \donnansource, reported in \citet{Donnan2022}, for which we find  a tentative $2.6\sigma$ detection at $850\,\rm\mu m$, and show that, if this emission is real and associated with this source, it  would imply a lower-redshift solution around $z\sim5$. Finally, we examine all the available long-wavelength (mid-IR to millimeter) observations around the  $z\approx12$ candidate known as \finkelsteinsource\ (\citealt{Finkelstein2022}), finding no evidence of continuum emission.

This Letter is organized as follows:
\S\ref{secc:observations} describes the new observations and the ancillary datasets. In \S\ref{secc:sed_fitting} we describe the  SED fitting methodology and the best-fit SED  fitting for \zavalasource\ along with the inferred physical properties. Then, \S\ref{secc:z16_source} introduces our search for potential contamination from  other DSFGs in   samples of $z>12$ LBGs candidates in the CEERS field including \donnansource\ and \finkelsteinsource. Finally, our conclusions are summarized in \S\ref{secc:conclusions}.

In this Letter, we assume $H_0=67.3\,\rm km\,s^{-1}\,Mpc^{-1}$, $\Omega_\lambda=0.68$, and $\Omega_{\rm M}=0.32$ (\citealt{Planck2016a}).

\section{Observations}\label{secc:observations}
\subsection{NOEMA observations}
We obtained NOEMA continuum observations on a sample of 19 DSFG candidates in the Extended Groth Strip (EGS) field in preparation for CEERS JWST data, as part of the NOEMA Program W20CK (PIs: Buat \& Zavala). The targets were selected from the original sample reported in \cite{Zavala2017a,Zavala2018a} based on deep observations at both 450 and 850$\,\mu\rm m$ obtained with the SCUBA-2 camera on the James Clerk Maxwell Telescope (JCMT). Here, we only focus on CEERS-DSFG-1 (known as 850.027 in \citealt{Zavala2017a,Zavala2018a}). The rest of the observations, along with a detailed description of the sample selection, will be presented elsewhere (Ciesla et al. in preparation).

NOEMA observations were performed using the wideband correlator {\it Polyfix}  covering the frequency ranges $252.5-260\,\rm GHz$ (with the lower side band) and $268-275.5\,\rm GHz$ (with the upper side band). The on-source integration time varies from $\sim10$ to $\sim50\,$minutes and was determined based on the 850$\,\mu\rm m$ flux densities of each target. For the main target of this Letter, \zavalasource, the on-source integration time was around $25\,$minutes. Calibration and imaging of the {\it uv} visibilities were then performed with {\sc gildas}\footnote{\url{www.iram.fr/IRAMFR/GILDAS}}, producing continuum maps with $0\farcs15\times0\farcs15$ pixels centered at $270\,\rm GHz$. For  \zavalasource, the achieved RMS is measured to be $\sigma_{\rm 1.1\,mm}=0.10\,$mJy\,beam$^{-1}$ and the beam size  $1\farcs35\times0\farcs85$. The continuum flux density at 1.1\,mm was extracted using an aperture of $1.5\times$ the beam size to recover any potential extended emission resolved by the beam.

Our NOEMA observations did not explicitly target the other two sources we include in this Letter, \donnansource\ or \finkelsteinsource, although the former is covered in a low sensitivity, outlying part of the primary beam of the observations of \zavalasource.  We discuss this further in \S~\ref{secc:z16_source} below.

\subsection{CEERS data}
JWST/NIRCam observations were conducted as part of the CEERS (Finkelstein et al., in prep) Survey program, one of the early release science surveys (\citealt{Finkelstein2017a}). Here, we only use data from CEERS pointing \#2, which covers all three objects we study in seven  filters: F115W, F150W,  F200W, F277W, F356W, F410M, and F444W.
After a three-dither pattern, the total exposure  time was typically 47\,minutes per filter, with the exception of F115W, whose integration time is longer (see details in \citealt{Finkelstein2022} and Finkelstein in prep.).

We performed a detailed reduction as described in \citet{Bagley2022} and \citet{Finkelstein2022b}. What follows is a brief summary of the main steps, and we refer the reader to these two papers for more details. We used version 1.7.2 of the \textit{JWST} Calibration Pipeline\footnote{\url{jwst-
pipeline.readthedocs.io}}, with custom modifications. Raw images were processed through Stages 1 and 2 of the pipeline, which apply detector-level corrections, flat-fielding, and photometric flux calibration. We also applied a custom step to measure and remove $1/f$ noise. We align the F200W images to an \textit{HST}/WFC3 F160W reference catalog created from $0\farcs03\,\rm pixel^{-1}$ mosaics in the EGS field with astrometry tied to Gaia-EDR3 \citep[see][for more details about the methodology]{koekemoer2011}. We then aligned each NIRCam filter to F200W, achieving a median astrometric offset $\lesssim0\farcs005$. Our steps represent an initial reduction that will be iteratively improved with updates to the Calibration Pipeline and reference files.

The flux extraction was done following  Finkelstein et al. (in prep.). Briefly, we use a multiwavelength photometric catalog created with Source Extractor \citep{Bertin1996a}, that was created with a sum of F277W$+$F356W as the detection image, with colors measured in small Kron apertures on images PSF-matched to F444W.  Total fluxes were estimated following an aperture correction based on a ratio between a large Kron (MAG\_AUTO) flux and the small Kron flux in the F444W image, with an additional correction for missing light in the large aperture based on simulations.  Finally, a systematic offset of 1-5\% was applied based on comparing the colors of best-fitting model templates to the photometry for $\sim$800 spectroscopically confirmed galaxies.

\subsection{Other ancillary data}

Photometric constraints at 450 and 850$\,\mu\rm m$ were obtained from \citet{Zavala2017a}, who reported deep observations with a central depth of $\sigma_{450\,\mu\rm m} = 1.2\,$mJy\,beam$^{-1}$ and $\sigma_{850\,\mu\rm m} = 0.2\,$mJy\,beam$^{-1}$, respectively, with a beam size of  $\theta_{450\,\mu\rm m}\approx8''$ and $\theta_{850\,\mu\rm m}\approx14\farcs5$.

We also make use of {\it Spitzer} IRAC 8$\,\mu$m (\citealt{Barro2011a}) and MIPS 24$\,\mu$m \citep{Magnelli2009} observations, as well as {\it Herschel} photometry from PACS (at 100 and 160$\,\mu\rm m$; \citealt{Lutz2011a}) and SPIRE (at 250, 350, and 500$\,\mu\rm m$; \citealt{oliver2012a}). Note, however, that the sources studied here are not detected in the {\it Spitzer} or {\it Herschel} maps and so we adopt only upper limits.  In addition, we use a 3\,GHz mosaic of the EGS field (Dickinson, private communication) obtained using observations from the Karl G. Jansky Very Large Array (VLA) as part of the program 21B-292 (PI: M. Dickinson). It reaches a sensitivity of $1.5\,\rm \mu Jy\,beam^{-1}$ and angular resolution of $2.3\times2.3\,\rm arcsec$.

The photometry extracted from these observations is summarized in Table \ref{tab:tab2}.

\section{A JWST/NIRCam dropout: A DSFG at redshift five}\label{secc:sed_fitting}

\begin{figure}
\centering
\includegraphics[width=0.85\columnwidth]{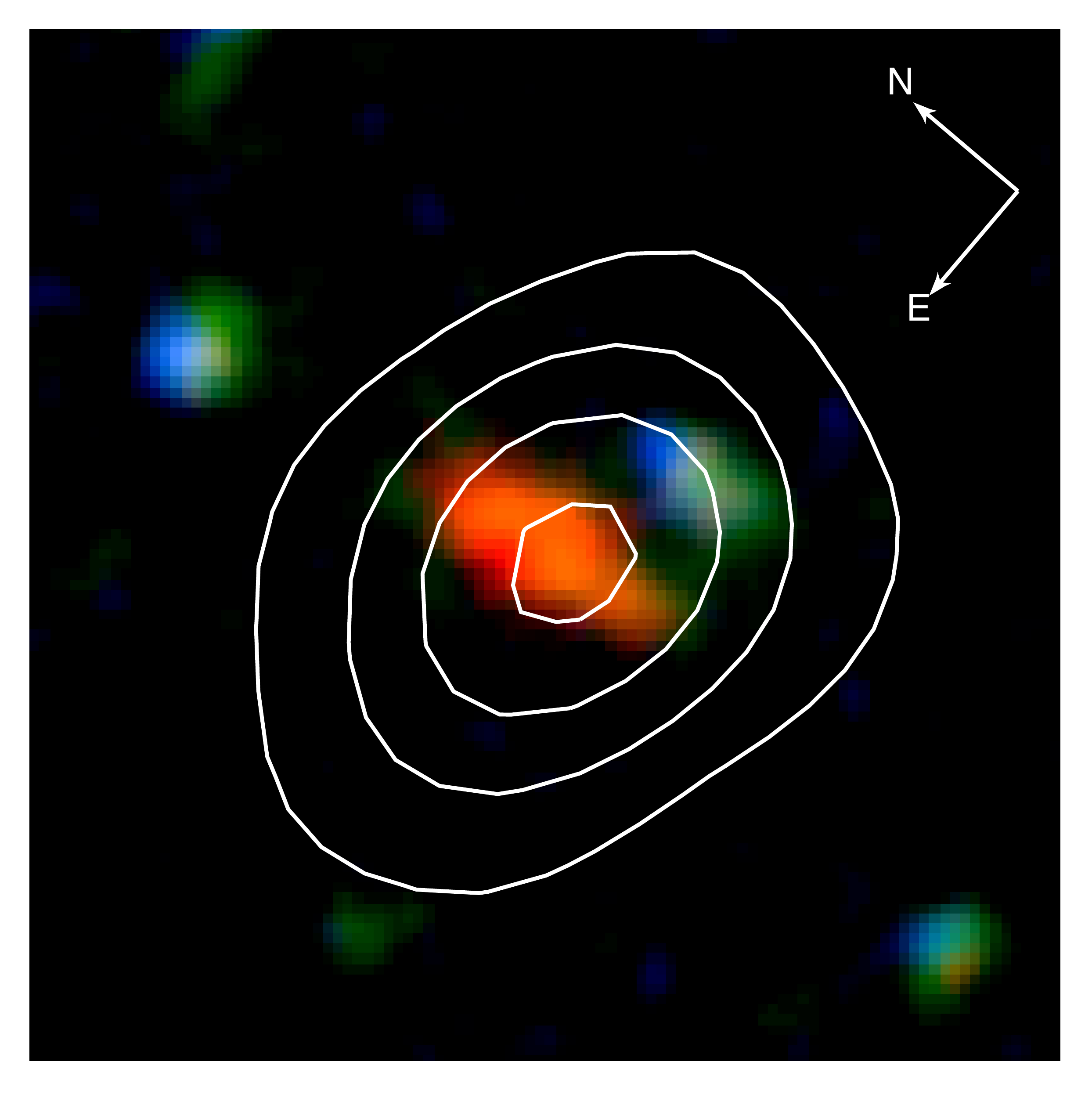}
\caption{A $3\farcs0\times3\farcs0$ composite image centered at the position of \zavalasource; the JWST/NIRCam F115W observations are in blue, F277W in green, and F444W in red (the data have been smoothed to roughly match the F444W resolution for better visualization). The 1.1\,mm NOEMA signal-to-noise ratio levels starting at 2.5$\sigma$ to 10$\sigma$ (in steps of 2.5$\sigma$) are represented by the white contours, clearly indicating that the dust thermal emission detected at submillimeter/millimeter wavelengths corresponds to the position of \zavalasource.
\label{fig:RGB_postage}}
\end{figure}

\begin{deluxetable}{ccccccccc}[h]
\vspace{2mm}
\tabletypesize{\small}
\tablecaption{Measured Photometry of \zavalasource}
\tablewidth{\textwidth}
\tablehead{
\colhead{Instrument/Filter} & \colhead{Wavelength} & \colhead{Flux Density}\\
}
\startdata
NIRCam/F115W & 1.15\,\um & $-$8$\pm$11\,nJy \\
NIRCam/F150W & 1.50\,\um & 18$\pm$13\,nJy \\
NIRCam/F200W & 2.00\,\um & 41$\pm$13\,nJy \\
NIRCam/F277W & 2.77\,\um & 137$\pm$8\,nJy \\
NIRCam/F356W & 3.56\,\um & 259$\pm$8\,nJy  \\
NIRCam/F410M & 4.10\,\um & 420$\pm$15\,nJy \\
NIRCam/F444W & 4.44\,\um & 438$\pm$12\,nJy \\
{\sc PACS}/100\,\um & 100\,\um & 0.11$\pm$0.51\,mJy \\
{\sc PACS}/160\,\um & 160\,\um & 0.1$\pm$3.5\,mJy \\
{\sc SPIRE}/250\,\um & 250\,\um & $-$1.1$\pm$5.8\,mJy \\
{\sc SPIRE}/350\,\um & 350\,\um & $-$4.5$\pm$6.3\,mJy \\
{\sc Scuba-2}/450\,\um & 450\,\um & $-$2.5$\pm$1.7\,mJy \\
{\sc SPIRE}/500\,\um & 500\,\um & $-$1.0$\pm$6.8\,mJy \\
{\sc Scuba-2}/850\,\um & 850\,\um & 2.25$\pm$0.36\,mJy \\
{\sc NOEMA}/1.1\,mm & 1.1\,mm & 1.92$\pm$0.11\,mJy \\
\enddata
\tablecomments{AB magnitudes can be derived via $-2.5\,\rm log_{10}(f_\nu[\rm nJy])+31.4$. \zavalasource\ is formally not detected in F115W, all of the {\it Herschel} bands from 100 to 500\,$\mu$m, and SCUBA-2 450\,$\mu$m.}
\label{tab:tab2}
\end{deluxetable}

\begin{figure*}
\centering
\includegraphics[width=\textwidth]{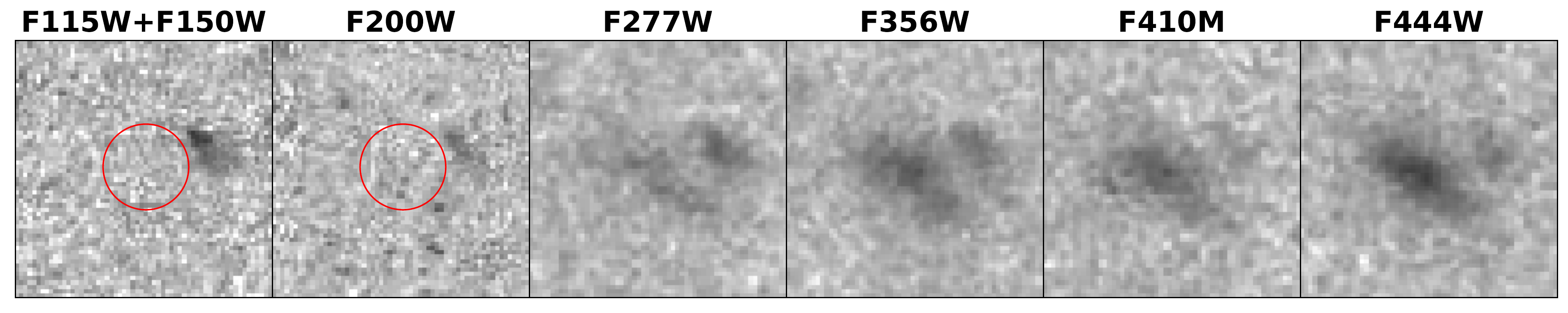}
\caption{1$\farcs$8$\times$1$\farcs$8 cutouts around \zavalasource\ from the CEERS JWST/NIRCam bands. The source is undetected in F115W, F150W, and F200W (the source's position is indicated with the red circle in the stacked F115W$+$F150W and F200W images, the dropout bands). The galaxy is well detected in redder filters with a red spectral shape. All images follow the same color code with a maximum value equal to $15\times$ the sky RMS and a minimum value of $-1.5\sigma$ and have the same orientation as Figure \ref{fig:RGB_postage}.
\label{fig:postages}}
\end{figure*}

The subarcsecond positional accuracy of the NOEMA observations allows us to directly identify the submillimeter-selected galaxy, \zavalasource, in the JWST/NIRCam observations (see Figure \ref{fig:RGB_postage}) without any ambiguity.
Interestingly, \zavalasource\ is well detected in F200W and redder bands but abruptly drops out in
F150W and F115W and in all the {\it HST} filters, as can be seen in Figure \ref{fig:postages}. The dropout nature of this source satisfies some of the color  criteria to identify  $z>10$ galaxy candidates. Indeed, it satisfies the criterion
of $m_{150\rm W}-m_{200\rm W}>0.8$ used in
\citet{Yan2022}, and some (but not all) of the criteria used in \citet{Donnan2022}, with a $2\sigma$ non-detection in F115W and F150W, and $>3\sigma$ detections in redder filters (see Figure \ref{fig:postages} and Table \ref{tab:tab2}).
However, the identification of this source as a DSFG  calls into question  such a  very high-redshift scenario, given that the highest-redshift dust continuum detections ever reported are at $z\sim7-8$ \citep{Laporte2017a,Strandet2017a,Marrone2018a,Tamura2019a,Inami2022}. Moreover, the (sub)millimeter emission would imply an extreme  infrared (IR) luminosity in excess of $\sim10^{13}\, L_\odot$ and a large dust mass in tension with current models. This is thoroughly discussed in  Appendix \ref{app:k-correction}, where we show that relatively bright (sub)millimeter sources are unlikely to lie at $z>10$.

Here we conduct a more thorough investigation as to the possible redshift of \zavalasource\ using JWST constraints alone, (sub-)millimeter constraints alone, and a combination of both JWST and long-wavelength millimeter data.
The results are highly dependent on the available photometric constraints and the inferred redshifts differ significantly, as discussed below.

\subsection{SED fitting procedure and redshift constraints}
\subsubsection{{\sc EAZY}}
\begin{figure}
\hspace{-0.6cm}
\includegraphics[width=0.5\textwidth]{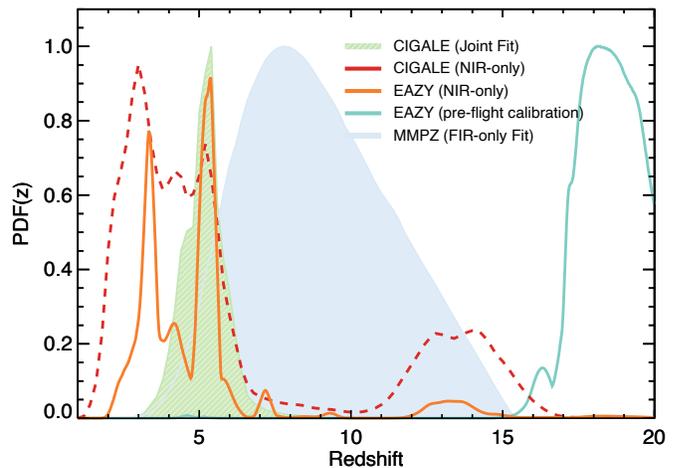}
\caption{Normalized redshift probability density distributions for \zavalasource\ from the different SED fittings.
The broad solid light blue distribution centered at $z\sim7.7$ comes from the (sub)millimeter wavelength constraints using {\sc MMPz} \citep{Casey2020a}. The fits to the JWST photometry alone from EAZY (orange) and CIGALE (red) are also very uncertain and show non-negligible probabilities at several very different redshift ranges. On the other hand, the joint {\sc CIGALE} fit to the NIRCam and the (sub)millimeter data results in a narrow distribution, constraining the redshift of \zavalasource\ at $z=5.09^{+0.62}_{-0.72}$.
Finally, to illustrate the impact of pre-flight NIRCam calibrations on the redshifts constraints, we also show the results from an EAZY fit to previous reductions of the NIRCam imaging (cyan line), which pushed the redshift constraints at $z>15$.
\label{fig:zPDF}}
\end{figure}

\begin{figure*}
\centering
\includegraphics[width=\textwidth]{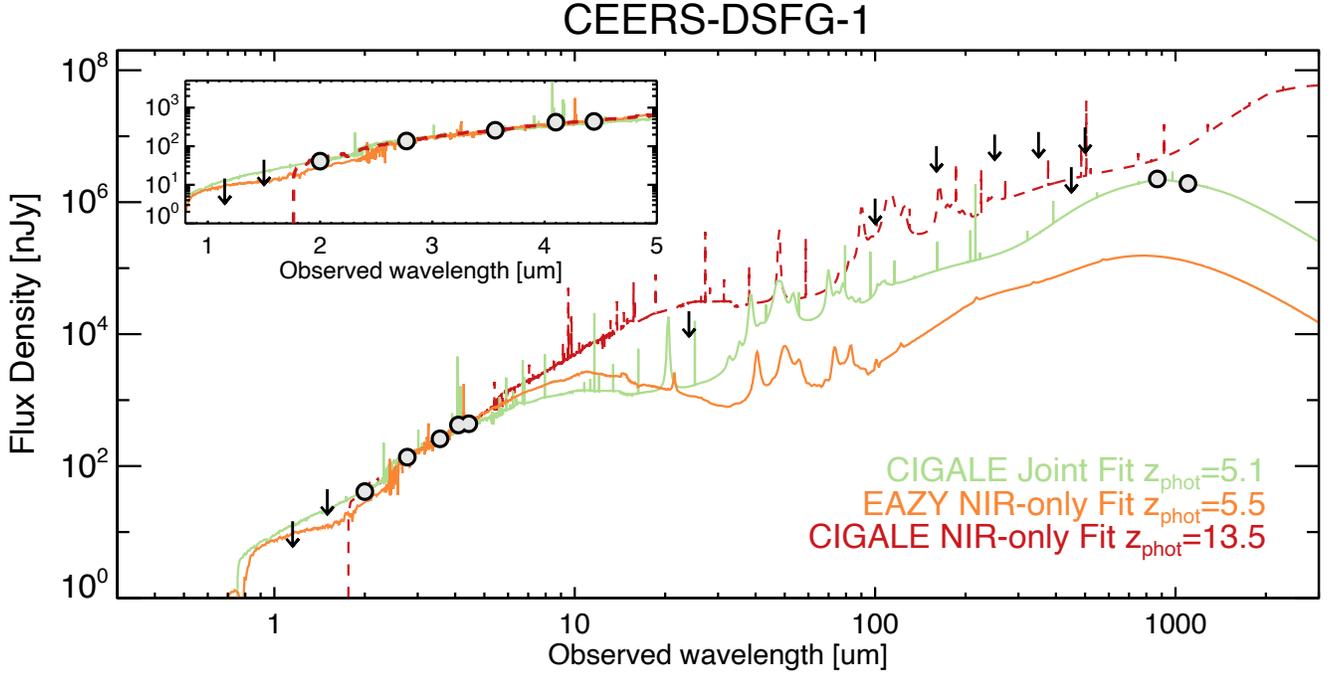}
\caption{The full near-infrared through millimeter spectral energy distribution of \zavalasource\ overplotted with several of the best-fit SEDs described in \S~\ref{secc:sed_fitting} along with the current photometric constraints. The detections used in the fits are represented by the solid circles, while  $2\sigma$ upper limits are illustrated by the downward arrows (note that error bars are smaller than the symbols). We show the
EAZY fit to near-IR data only at $z=5.5$ in orange,
{\sc CIGALE}  to near-IR data only at
$z=13.5$ in red, and the fiducial {\sc CIGALE} joint fit at  $z=5.1$ in light green.
While the three SEDs satisfactorily reproduce the JWST photometry (see zoom-in inset plot), only the joint fit at $z=5.1$ reproduces the (sub)millimeter fluxes.
\label{fig:fullsed}}
\end{figure*}

We first fit the SED of \zavalasource\ to JWST/NIRCam photometry alone using the {\sc eazy} (\citealt{Brammer2008a}) SED fitting code. The fitting was performed in an identical fashion as in \citet{Finkelstein2022}. To summarize, EAZY makes use of a user-supplied template set to generate linear combinations of stellar populations that fit the data and generate redshift probability distributions.  The template set used in our case includes the ``tweak\_fsps\_QSF\_12\_v3'' set of 12 templates as well as 6 additional templates that span bluer colors (\citealt{Larson2022}).
As shown in Figure \ref{fig:zPDF}, the redshift probability density distribution from EAZY shows two significant peaks at $z\sim3$ and $z\sim5$, and a non-negligible probability ($\sim6\%$) at $z\approx12-14$. To put these fits in context with the (sub)millimeter data, we show in Figure \ref{fig:fullsed} the best-fit SED from EAZY at $z=5.5$, corresponding to the redshift with the maximum probability.

\subsubsection{{\sc CIGALE}}\label{secc:cigale}

We also fit the photometry using {\sc CIGALE} \citep{Burgarella2005,Noll2009,Boquien2019}  assuming a delayed star formation history (SFH): SFR(t) $\propto t \exp(-t/\tau)$ with stellar models from  \citet{Bruzual_Charlot2003} (BC03). A \citet{Calzetti2000} law is also adopted for the dust attenuation of the stellar continuum. On the other hand, the nebular emission (continuum and lines) is attenuated with a screen model and an SMC extinction curve (\citealt{Pei1992a}).  Finally, the dust emission reemitted in IR is modeled with \citet{Draine2014} models.

Including only JWST/NIRCAM photometry in the fit results in a similar redshift distribution as the one obtained with EAZY, with significant probability at $z\approx3-5$ and a moderate probability of $\approx22\%$ of being at $z>10$. To illustrate how well the high-redshift solutions fit the available data, we include the  best-fit SED at $z=13.5$ in Figure \ref{fig:fullsed}.

In addition, we fit the JWST data along with SCUBA-2 and NOEMA detections ({\it Herschel} upper limits  were not  included in the fit) using the same CIGALE configuration described above. The addition of the long-wavelength data significantly impacts the results, narrowing down the redshift probability distribution of \zavalasource\ (see Figure \ref{fig:zPDF}). The best-fit photometric redshift when using all the available photometric constraints is $z=5.09^{+0.62}_{-0.72}$, where the error bars encompass the 68\% confidence interval. As shown in Figure \ref{fig:fullsed}, the fitted SED from this analysis is in good agreement with all the available photometric constraints, including upper limits.

\subsubsection{MMPz}

Finally, though the long-wavelength data on \zavalasource\ are somewhat limited, we are able to calculate an independent photometric redshift for the source based on long-wavelength data alone using the {\sc MMPz} package \citep{Casey2020a}.  {\sc MMPz} presumes that sources with significant (sub)millimeter emission follow an empirically measured relationship between the rest-frame peak wavelength of emission, $\lambda_{\rm peak}$, which is inversely proportional to the characteristic luminosity-weighted dust temperature of the ISM, and the total emergent IR luminosity, $L_{\rm IR}$.  This $L_{\rm IR}-\lambda_{\rm peak}$ relation is fairly well constrained out to $z\sim5$ \citep{casey2018a,drew2022a} where more intrinsically luminous sources have warmer temperatures. {\sc MMPz} generates a redshift probability distribution by computing the $L_{\rm IR}$ and $\lambda_{\rm peak}$ at all possible redshifts, and contrasts that against the empirical distribution of measured SEDs.  By design, redshift solutions found using {\sc MMPz} are very broad (due to the degeneracy between ISM dust temperature, constrained via $\lambda_{\rm peak}$, and redshift). The best-fit redshift generated from the long wavelength data alone (including the only two detections and all the non-detections) is most consistent with the joint CIGALE fit, but shifted to higher values with a best-fit redshift of $z=7.77^{+2.55}_{-1.69}$ (see Figure \ref{fig:zPDF}).

\subsubsection{The moral of the story}
From the above analysis, it is clear that a single color (i.e. drop-out) selection criteria to identify high-redshift ($z\gtrsim10$) candidates might include contamination from lower-redshift sources, such as \zavalasource.
This contamination could be more severe in studies using pre-flight calibrations since they render the colors of some galaxies more akin to those expected for very high-redshift systems (see discussion by \citealt{Adams2022a}). This is clearly illustrated in Figure \ref{fig:zPDF}, where we have also included the photometric redshift constraints from EAZY using a pre-flight calibration. The fit suggests a very high redshift of $z=18.2^{+1.2}_{-0.7}$ with an almost negligible probability at $z<15$.
Careful selection criteria (with several conditions) are thus necessary to produce cleaner samples of high-redshift galaxies.
\citet{Finkelstein2022b} and \citet{Harikane2022}, for example, implemented a further criterion based on the significance of the high-redshift solution against secondary lower-redshift solutions (defined by the difference between the $\chi^2$ values of the high-redshift and low-redshift solutions) to select robust candidates (see also \citealt{Donnan2022}). Similarly, other studies used a two-color criterion to minimize contaminants (e.g. \citealt{Adams2022a,Atek2022,Castellano2022,Harikane2022}) that would have prevented the selection of \zavalasource\ as a very high-redshift candidate given its red colors at longer wavelengths (e.g. $m_{277\rm W}-m_{444\rm W}>1.26$).
Note, however, that despite these extra selection criteria, lower-redshift systems might still masquerade (and be misidentified) as very high-redshift galaxies as discussed in \S\ref{secc:z16_source}.

\subsection{On the physical properties of \zavalasource}
As mentioned above, the joint fit of CIGALE using the JWST/NIRCam and the (sub)millimeter data provide tight constraints on the redshift of our target and its physical properties. Hence, here we adopt the these results as our fiducial values. The inferred physical properties are summarized in Table \ref{tab:properties} and discussed below.

\begin{deluxetable}{lc}
\vspace{2mm}
\tabletypesize{\small}
\tablecaption{Properties of \zavalasource}
\tablewidth{\textwidth}
\tablehead{\multicolumn{1}{c}{Property} & \multicolumn{1}{c}{Value}}
\startdata
Source ID&CEERSJ141938.19$+$525613.9\\
R.A.~(J2000 [deg])&214.9091152\\
Decl.~(J2000 [deg])&52.9371977\\
$z_{\rm CIGALE}$ & $5.09^{+0.62}_{-0.72}$\\
$M_{\star}~(M_{\odot}$) & $(2.1 \pm 0.8)\times10^{10}$ \\
$L_{\rm IR}~(L_{\odot}$) & $(1.1 \pm 0.3)\times10^{12}$ \\
SFR~($M_{\odot}$ yr$^{-1}$) & $110 \pm  30$\\
sSFR~(Gyr$^{-1}$) & $5.2 \pm  2.5$\\
$E(B-V)$~(mag) & $1.6\pm 0.1$ \\
Age~(Myr) & 490$\pm 240$\\
Mass-weighted age~(Myr) & 170$\pm 90$\\
\enddata
\tablecomments{The redshift and the listed physical properties were derived from  the joint fit of CIGALE using the JWST/NIRCam data and the available (sub)millimeter constraints.}
\label{tab:properties}
\vspace{-8mm}
\end{deluxetable}

Assuming the best-fit redshift of  $z=5.1$, the stellar mass of \zavalasource\ is constrained to be
$(2.1 \pm 0.8)\times10^{10} M_\sun$. This is a factor of $\sim4$ smaller than the average mass of DSFGs detected by single-dish telescopes (e.g. \citealt{daCunha2015a}), but is aligned with expectations since our source was selected from one of the deepest SCUBA-2 surveys and has a fainter 850\,\um\ flux density than  typical galaxies identified in shallower single-dish telescope surveys. Indeed, the stellar mass of our target is in better agreement with other SCUBA-2 galaxies identified in this field, which have an average stellar mass of $\approx5.6\times10^{10} M_\sun$ (\citealt{Cardona-Torres2022a}), and with the masses derived for galaxies identified in recent deeper Atacama Large Millimeter/submillimeter Array (ALMA) surveys (e.g. \citealt{Gomez-Guijarro2022}; see also \citealt{Khusanova2021}).
Similarly, the SFR of \zavalasource\ of  $\rm 110\pm  30\, M_\sun~yr^{-1}$ (averaged over the past 10\,Myr) lie between those from SMGs and fainter DSFGs identified in deeper ALMA observations (\citealt{daCunha2015a,Zavala2018b,Aravena2020a,Casey2021a,Khusanova2021,Gomez-Guijarro2022}).
These properties imply a specific star formation rate of $\rm sSFR=5.2\pm2.5\,\rm Gyr^{-1}$, meaning that \zavalasource\ lies on the main-sequence of star forming galaxies, similar to the so-called population of ``{\it HST}-dark" galaxies\footnote{\zavalasource\ is, by definition, an ``HST-dark''  galaxy.} (e.g. \citealt{Wang2019a}).

At $z = 5.1$, the NIRCam photometry samples rest-frame wavelengths from 0.2 to 0.7$\,\mu$m, allowing us to constrain the stellar dust attenuation.
The red spectral shape in the NIRCam bands implies a strong dust attenuation (as typically found for this kind of galaxies; e.g. \citealt{Simpson2017}) with $E(B-V)=1.6\pm 0.04$, which results in a dust luminosity of $(1.1 \pm 0.3)\times10^{12}\, L_\sun$.

\section{Searching for DSFG contaminants in high-redshift LBG candidates identified with JWST}\label{secc:z16_source}

The SCUBA-2 observations from \citet{Zavala2017a} partially overlap with the CEERS NIRCam survey and thus can be used to look for dust continuum emission around  $z>10$ candidates in the field. Here we focus on two recently reported high-redshift candidates: \donnansource\ reported to be at $z\approx16.7$  \citep{Donnan2022}  and \finkelsteinsource\ at $z\approx11.8$ (\citealt{Finkelstein2022}).

\subsection{A deeper look into \donnansource}
\begin{figure}
\centering
\includegraphics[width=0.85\columnwidth]{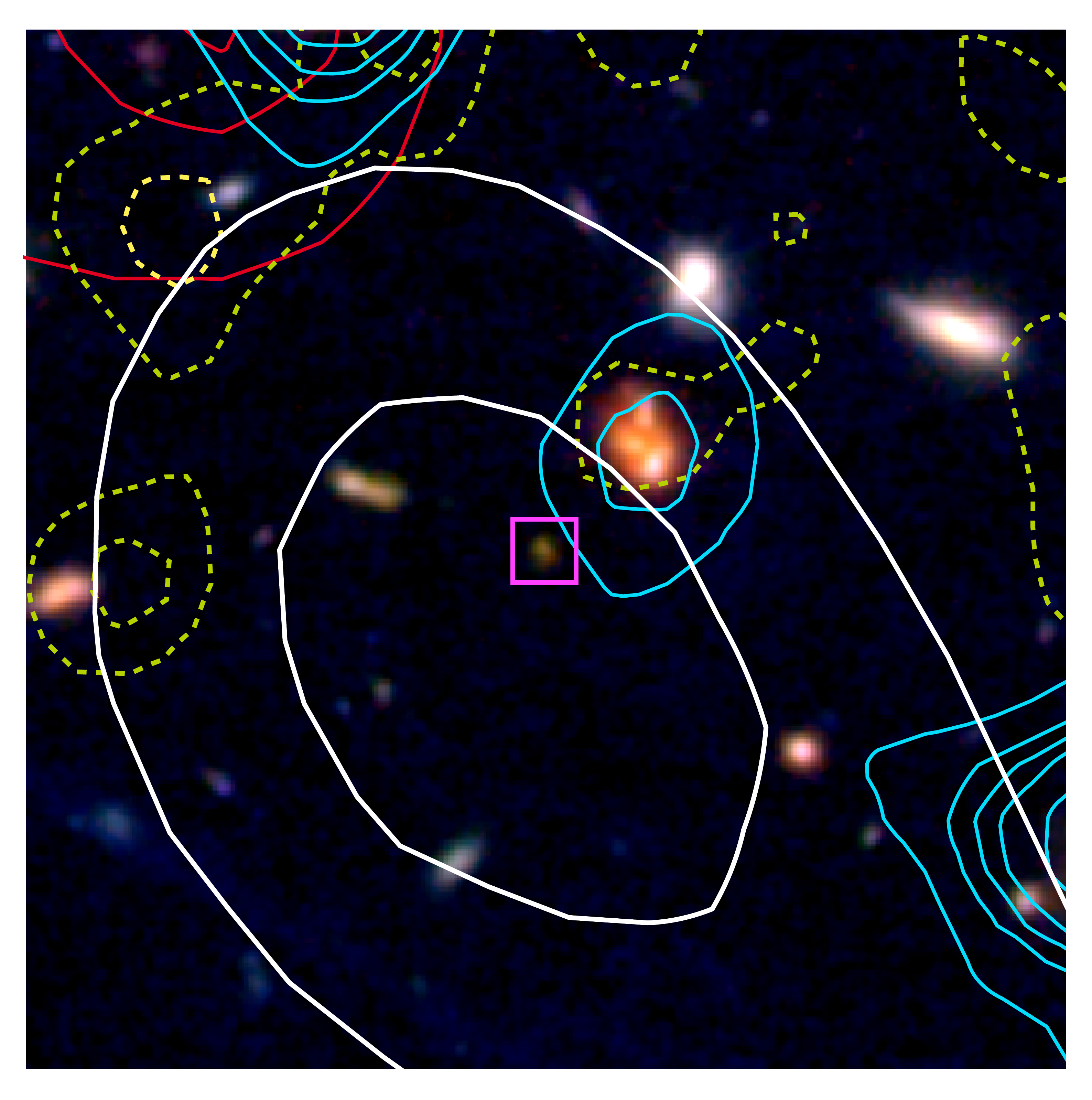}
\caption{A $12\farcs0\times12\farcs0$ composite image (blue: F115W; green: F277W; red: F444W) centered at the position of the $z=16.7$ candidate from \citet{Donnan2022} (magenta square).
The deep SCUBA-2 $850\,\mu\rm m$ map from \cite{Zavala2017a} shows a $\sim2.6\sigma$ positive emission at this position (white contours indicate $1\sigma$ and $2\sigma$ levels). We also plot signal-to-noise contours at $1\sigma$, $2\sigma$, $2.5\sigma$, $3\sigma$, and $4\sigma$ from {\it Spitzer} $8\,\mu\rm m$ and $24\,\mu\rm m$ as cyan and red lines, respectively, and from VLA 3\,GHz as yellow dashed lines.
From the lack of clear detections, we can rule out an obvious alternate radio or 24\,\um\ counterpart for the 850\,\um\ emission at $z\lesssim3$, while we cannot rule out galaxies at $z\gtrsim3$ since they are usually undetected at these wavelengths.
The marginal 8\,\um\ emitter on the top right (west) of \donnansource\ is a potential counterpart, but the current data are unconstraining.
\label{fig:edinburgh_source}}
\end{figure}

A $2.6\sigma$ tentative detection around the position of \donnansource\ (\citealt{Donnan2022}; R.A.=214.91450, decl.=52.943033) was found in the $850\,\mu\rm m$ SCUBA-2 map with a flux density of 0.65$\pm$0.26\,mJy (see Figure \ref{fig:edinburgh_source}).
Unfortunately, this source was not
formally targeted by our NOEMA observations and, although it is only $26''$ away from \zavalasource\ and within the coverage of the NOEMA map described above,
it lies on the edge of the map, where the sensitivity is very low (with a primary beam response of $\lesssim0.1$, implying an RMS of $\sigma_{\rm 1.1\,mm}\gtrsim1\,$mJy\,beam$^{-1}$).

\subsubsection{Caveats of a Marginal SCUBA-2 Detection}

We emphasize that there are two primary reasons why this marginal detection may not conclusively imply that \donnansource\ is a significant thermal dust emitter.
The first concern is the significance of the signal itself and the possibility of being spurious. At 2.6$\sigma$, simulations of blind detections, single-dish submillimeter sources indicate  false-positive rates as high as
$\sim30-40$\,\%\ \citep{Casey2013a,Casey2014a}.
These rates of false positives are estimated by both searching SCUBA-2 maps for negative significance peaks at $-$2.6$\sigma$ as well as conducting source injection tests on SCUBA-2 jackknife maps \citep[following the same methodology as][ see their Figure 7]{Casey2013a}. To complement these results, we test the reliability of these low signal-to-noise ratio peaks by creating a catalog of $2.5\sigma$ to $3.0\sigma$ SCUBA-2 sources, and searching for counterparts in the deep VLA 3\,GHz map  (Dickinson, private communication). We find clear associations for at least $50\%$ of the SCUBA-2 sources\footnote{Given the surface density of radio and the SCUBA-2 sources, the probability of chance alignment is $<5\%$}, implying a $\sim50\%$ fidelity rate. A similar result is obtained using the 24\,\um\ map. Note, however, that this reliability fraction of $50\%$ should be considered a lower limit since it is well-known that a significant fraction (as high as 30-40\%) of submm sources lack radio or mid-infrared counterparts (particularly those at $z>3$; \citealt{chapman2003,barger2007,Pope2006,Dye2008}).
We thus conclude that the $2.6\sigma$ SCUBA-2 signal around CEERS-93316 has a
$\gtrsim50\%$ probability of being real.

The second significant concern
is that even if the detection is real, the SCUBA-2 beamsize is large enough that the 850$\,\mu$m emission could arise from another galaxy at a close angular separation with \donnansource\ on the sky.
Figure~\ref{fig:edinburgh_source} shows the neighboring sources within the beamsize of the SCUBA-2 tentative detection, with contours overlaid for {\it Spitzer} 8\,\um\ emission, 24\um\ emission, and VLA 3\,GHz continuum.
Unfortunately, there is no secure emitter at these wavelengths to which we can definitively associate the 850$\,\mu$m  emission to unequivocally rule out association with \donnansource. Note that the lack of such a counterpart does not imply the tentative SCUBA-2 emission is spurious, since $z>3$ galaxies are usually undetected in these bands (this is indeed the case for \zavalasource). This lack of detection rather means that it is not implausible to associate the 850\,\um\ emission with \donnansource, although it also does not confirm the association. Another possible counterpart could be the 8\,\um\ emitter (with a $\sim2.5\sigma$ significance) to the northwest that has a photometric redshift of $z\sim5$, though it is farther from the signal-to-noise peak in the SCUBA-2 map than \donnansource.

At present, we lack sufficient data to clearly associate the emission with \donnansource\ or other neighboring sources. Follow-up interferometric observations would be necessary to provide both a confirmation of the emission and astrometric localization to \donnansource\ or to a neighboring source.
Nevertheless, given the remarkable properties of  \donnansource\ (being one of the highest-redshift candidates ever reported with a bright UV magnitude of $M_{\rm UV}=-21.7$),
below we explore the impact that the submillimeter tentative detection might have on its redshift solution {\it if} the dust emission is real {\it and} associated with it.

\subsubsection{Implications if Dust Emission is associated with \donnansource}

First, we consider what the implications would be if \donnansource\ had significant dust emission at its proposed redshift of $z=16.7$.
The observed 850\,\um\ emission would probe the rest-frame $\sim$50\,\um\ regime; in this scenario, the IR luminosity would be above $10^{12}\,L_\odot$ with a dust mass of $\sim$10$^{8}\,M_\odot$.  A system with such high dust mass found $\sim$230\,Myr after the Big Bang would surely be extraordinary, likely implausibly so  (e.g., \citealt{Dwek2014a}). This is further discussed in Appendix \ref{app:k-correction}, where we show the predicted IR luminosity and dust mass as a function of redshift for a hypothetical submillimeter detection with a flux density similar to that of the tentative emission discussed here.

We alternatively explore if a lower-redshift solution would be plausible given the JWST/NIRCam photometric constraints and the observed blue colors in these bands (which contrasts with those from \zavalasource).
To do that, we fit the JWST/NIRCam data\footnote{Note that since we performed our own data reduction and followed our own source extraction procedure designed to measure accurate colors, the NIRCam fluxes for \donnansource\ used in this Letter could differ from those in \citet{Donnan2022}. We list the adopted fluxes for the SED fitting in Appendix \ref{appendix}. } along with the tentative $850\,\rm\mu m$ flux density with CIGALE.

\begin{figure}
\hspace{-0.6cm}
\includegraphics[width=0.5\textwidth]{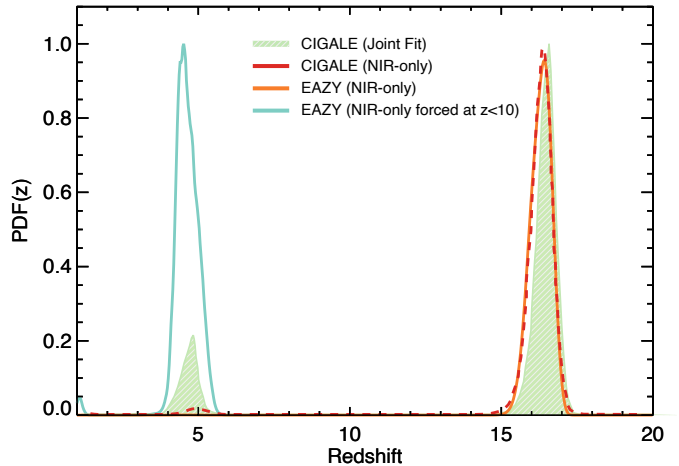}
\caption{Normalized redshift probability density distributions for \donnansource\ from the different SED fittings.
The EAZY and CIGALE fits to the JWST photometry alone are represented by the orange and red lines, respectively. The two fits suggest a high redshift around $\sim16.3$, in agreement with \citet{Donnan2022}. The addition of the tentative submillimeter emission to the CIGALE fit does not alter the main peak of the distribution, although it increases the probability at $z\sim4.8$ as illustrated by the green shaded area. This lower-redshift solution is in good agreement with the results from EAZY when imposing a maximum redshift at $z=10$  (cyan line). As thoroughly discussed in the main text, if the tentative 850$\,\mu\rm m$ emission is real and associated with \donnansource, our analysis suggests that this high-redshift galaxy candidate rather lies at a lower redshift around $z\sim4.8$.
\label{fig:zPDF_donnan}}
\end{figure}

\begin{figure*}
    \hspace{-0.1cm}
    \includegraphics[width=\textwidth]{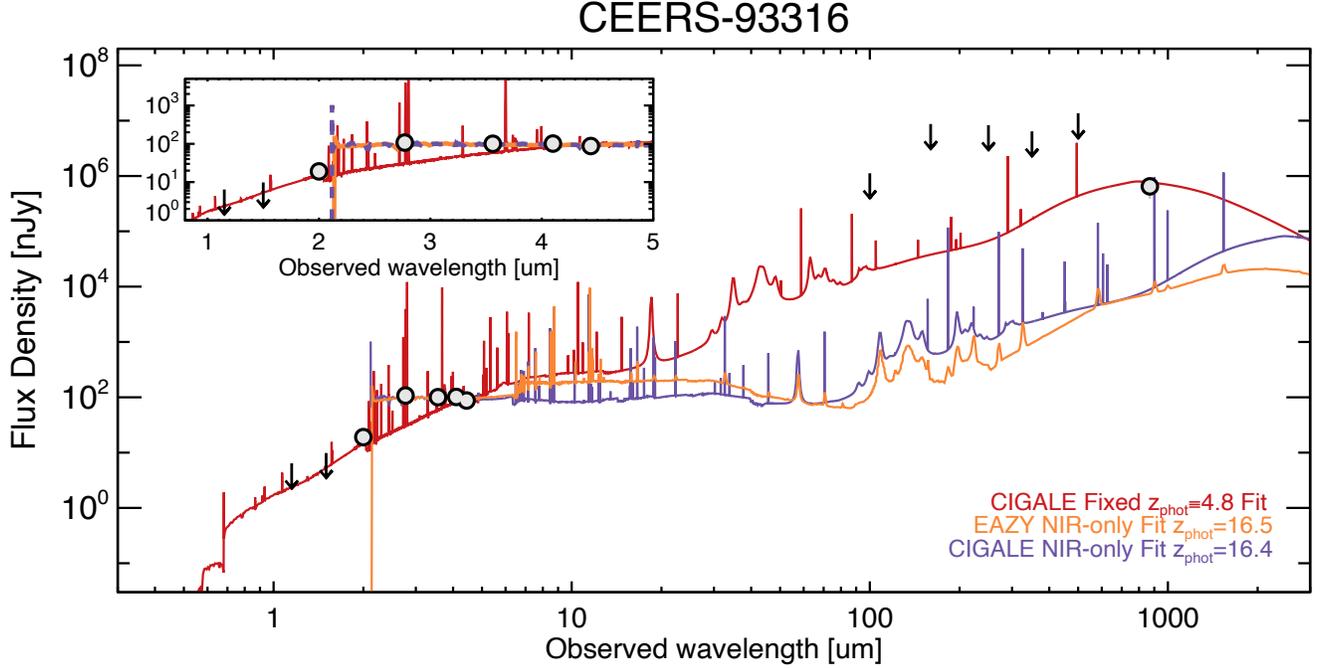}
    \caption{Best-fit spectral energy distributions for \donnansource. The orange and purple lines represent the $z\sim16.4$ EAZY and CIGALE fits to the near-IR data only, which predict a 850\,\um\ flux density significantly lower than the value implied by the tentative SCUBA-2 detection.
    On the other hand, the best-fit SED at $z=4.8$ from CIGALE
    (red line) can reproduce both the near-IR and the submillimeter data.
    In this case, the blue NIRCam colors are produced by the strong emission lines, as clearly seen in the inset plot. }
    \label{fig:fullsed_donnan}
\end{figure*}

While the redshift distribution from this fitting strongly favors a high-redshift solution in agreement with the \citet{Donnan2022} result (with $z_{\rm CIGALE}\approx 16.4$; see Figure \ref{fig:zPDF_donnan}), the best-fit SED does not satisfactorily reproduce the tentative submillimeter flux density that is under-estimated by more than an order of magnitude (see Figure~\ref{fig:fullsed_donnan}).
Interestingly, the redshift probability distribution does show a secondary peak at $z\sim4.8$, although with a low integrated probability of less than $3\%$ (see Figure \ref{fig:zPDF_donnan}). This peak is seen even without the inclusion of the long-wavelength emission and it is also seen in the redshift probability density distribution presented by  \citet[see their Figure A1]{Donnan2022}.
This lower redshift  clearly dominates the probability distribution of the EAZY fitting when imposing a maximum redshift\footnote{The $z=10$ threshold was chosen based on the discussion presented in Appendix \ref{app:k-correction}. Note that other works have also followed this strategy to  better assess the feasibility of low-redshift solutions (e.g. \citealt{Finkelstein2022b}).} of $z=10$, as  shown in the Figure \ref{fig:zPDF_donnan}.

To further explore the feasibility of this alternative redshift solution, we re-run CIGALE but fix the redshift to $z=4.8$.
The resulting SED is shown in Figure \ref{fig:fullsed_donnan} along with the best-fit $z\sim16$ SEDs from EAZY and CIGALE, for comparison.
In the low-redshift scenario, the strong break seen between F200W and F277W in \donnansource\
is attributable to strong [OIII] and H$\beta$  emission in the F277W band (see Figure \ref{fig:fullsed_donnan}). Similarly, the excess flux in F356W above the continuum, which produces a blue F356W-F410M color, would be attributable to H$\alpha$ emission.
The measured NIRCam photometry would thus require a young starburst  with strong nebular line emission to satisfy a $z\sim4.8$ solution, but this would be within the realm of expectation for an early-stage DSFG in formation at these redshifts.

The $z\approx4.8$ best-fit SED would imply an SFR averaged over 10 Myr of $20 \pm  10\,M_\sun\rm\,yr^{-1}$ and a stellar mass equal to $(1.4 \pm 0.5)\times10^{9}\,M_\sun$, with a dust attenuation (for both continuum and lines) of $E(B-V)=0.5\pm 0.1$ and a dust luminosity of $(1.7 \pm 0.8)\times10^{11}\,L_\sun$. These properties are in broad agreement with those derived for the relatively faint population of $z\sim7$ dusty galaxies in the REBELS survey (\citealt{Bouwens2020a,Inami2022}).
In addition, the line fluxes required to reproduce the given NIRCam photometry range from $\sim1\times10^{-18}-1\times10^{-17}$\,erg\,s$^{-1}$\,cm$^{-2}$, which are within the range of those predicted for \zavalasource.

While deep interferometric observations at millimeter wavelengths are required to confirm or refute dust continuum emission in this high-redshift candidate, here we show (see also Appendix \ref{app:k-correction}) that a $z\sim4.8$ scenario associated with a DSFG with strong nebular emission is plausible for \donnansource\ and highly likely if the submillimeter  emission is confirmed,  despite its blue near-IR colors that are usually associated with the emission of dust-free systems \citep[e.g.][and references therein]{Finkelstein2016}. If this lower-redshift solution is true, it would contrast with the low probability of being at $z<15$ inferred from the different redshift probability distributions shown in  Figure \ref{fig:zPDF_donnan} (see also \citealt{Donnan2022,Finkelstein2022b}). The reason for this low probability might be related to the low significance of the tentative SCUBA-2 detection in the case of CIGALE or with the adopted templates and the fitting approach for EAZY. Interestingly, \citet{Perez-Gonzalez2022}, who used a novel 2D fitting approach with a new set of SED templates, found a best-fit redshift of $4.59\pm0.03$ for this source (known as nircam2-2159 in \citealt{Perez-Gonzalez2022}) with a  low probability  of being at $z>10$.

\subsection{A deeper look into Maisie's Galaxy}

Given that the recently reported $z=11.8$ galaxy candidate from \citet{Finkelstein2022} lies close to the two galaxies described above ($\sim78''$ and $\sim65''$ away from \zavalasource\ and from \donnansource, respectively), we carefully examine the available long-wavelength observations to investigate any possible detection of dust emission.

\begin{figure}
\centering
\includegraphics[width=0.7\columnwidth]{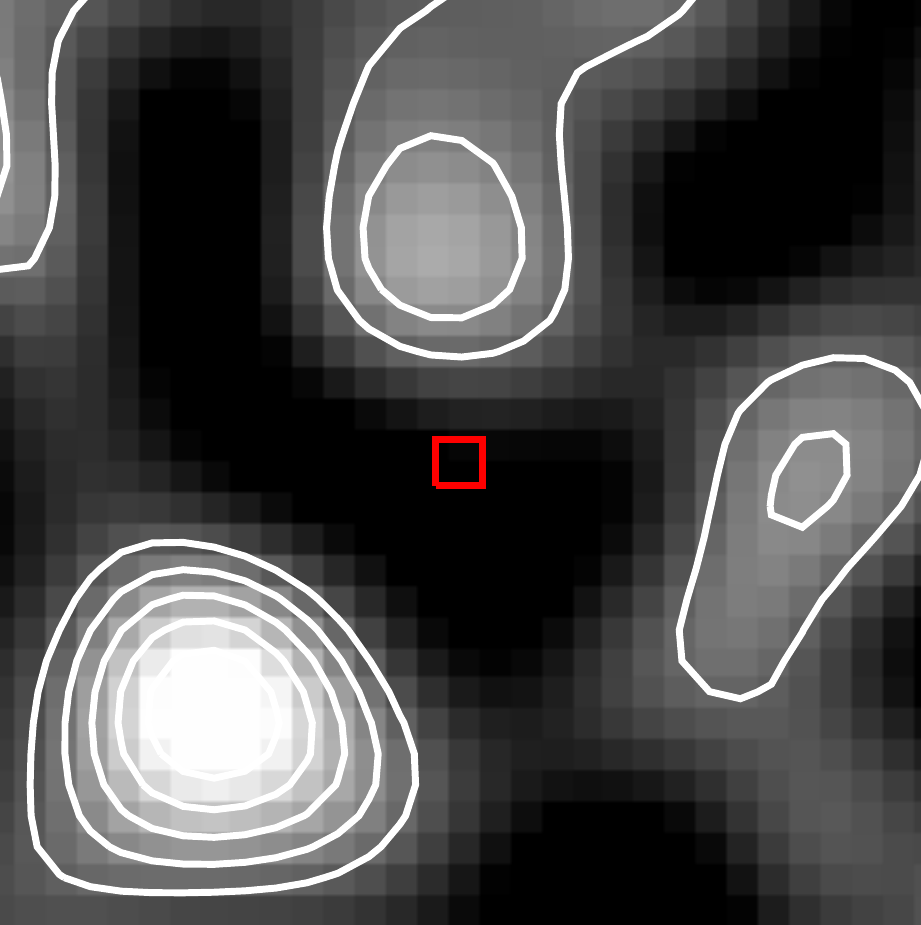}
\caption{SCUBA-2 $850\,\mu\rm m$ signal-to-noise ratio map ($60\farcs0\times60\farcs0$) centered at the position of Maisie's Galaxy (red square; \citealt{Finkelstein2022}). At the depth of the observations ($\sigma_{850\mu\rm m} = 0.46\,\rm mJy\,beam^{-1}$), no significant dust emission is detected (the flux density at the position of the source is measured to be $-0.4\pm0.25\,$mJy\,beam$^{-1}$). There is also no statistical significant emission in the available {\it Spitzer} (8 and $24\,\mu\rm m$), {\it Herschel} ($100-500\,\mu\rm m$), or SCUBA-2 ($450\,\mu\rm m$) maps.
\label{fig:maisies_source}}
\end{figure}

Because this source is not covered by our NOEMA observations, we started by looking at the deep SCUBA-2 850\,\um\ map (\citealt{Zavala2017a}).
As shown in Figure \ref{fig:maisies_source} no significant detection is found (with a measured flux density of $S_{\rm 850}=-0.40\pm0.25$\,mJy at the position of the source). We also search for significant emission in the {\it Spitzer} 8\,\um\ and 24\,\um\ maps, Herschel 100, 160, 250, 350, and 500\,\um\ imaging; and SCUBA-2 450\,\um\ observations, finding only non-detections. We thus conclude that a lower-redshift scenario for \finkelsteinsource\, in which the source is rather associated with a DSFG is unlikely. Given the lack of FIR-to-submillimeter detections, the best-fit SEDs and their associated redshift probability distributions for \finkelsteinsource\ would be similar to those reported in \citet{Finkelstein2022}. Hence, to avoid duplication, they are not included in this Letter.

\section{Conclusions}\label{secc:conclusions}

Using the available datasets from the JWST CEERS survey in combination with NOEMA and SCUBA-2 observations, we have demonstrated that DSFGs at $z\sim4-6$ can drop out in the bluest JWST/NIRCam filters while being well detected in the redder filters.  This kind of galaxies could even show a significant probability of being at high redshifts when performing SED fittings.
This is illustrated by studying the source \zavalasource, a 850\,\um-selected galaxy with robust interferometric observations at 1.1\,mm by NOEMA that is undetected in the F115W and F150W bands. A joint SED fitting analysis including the NIRCam constraints and the long-wavelength (sub)millimeter data implies a photometric redshift of $5.09^{+0.62}_{-0.72}$, with physical properties that resemble other DSFGs: $M_\star=(2.1 \pm 0.8)\times10^{10}\,M_\odot$; ${\rm SFR}=110\pm30\,M_\odot\rm\,yr^{-1}$; $L_{\rm dust}=(1.1 \pm 0.3)\times10^{12}\,\rm L_\sun$.
Hence, searches of $z>10$ LBGs that rely only on a  dropout selection
could introduce significant contaminants from lower-redshift systems.  This could be minimized by adopting multi-color selection criteria or by defining alternative conditions (such as a minimum redshift probability or $\chi^2$ goodness-of-fit; e.g. \citealt{Adams2022a,Castellano2022,Donnan2022,Finkelstein2022,Harikane2022}).

Taking advantage of the available submillimeter data in the field, we extended the search for dust continuum emission to two close $z>10$ LBG candidates recently reported, \donnansource\ at $z\approx16.7$ (\citealt{Donnan2022}) and \finkelsteinsource\ at $z\approx11.8$ (\citealt{Finkelstein2022}). We found a tentative $2.6\sigma$ detection at 850\,\um\ around the position of \donnansource.
A confirmation of this flux density measurement and a firm spatial association requires higher-resolution submillimeter imaging. This is particularly important given its high probability of  being spurious and the large beam size ($\approx 14.6''$) of the SCUBA-2 observations that encompass several galaxies.

While additional observations are required to corroborate this identification, we use this possible association to illustrate that $z\sim5$ DSFGs can also exhibit blue colors in the JWST/NIRCam bands when strong nebular emission lines are present (with line fluxes on the order of $\sim$10$^{-18}$--10$^{-17}$\,erg\,s$^{-1}$\,cm$^{-2}$), and  conclude that (sub)millimeter emission in samples of $z>10$ LBGs likely implies misidentifications of DSFGs at lower redshifts ($z\lesssim7$). Indeed,
if \donnansource\ is confirmed to be a dust emitter, our analysis suggests that it would rather lie at $z\sim5$.

This work has illustrated both the importance and potential of combining JWST observations with  submillimeter/millimeter data, a synergy that allows us to identify and characterize populations of galaxies that were previously unreachable, including both $z\gtrsim5$ DSFGs as well as ultra-high redshift $z>10$ LBGs.  In particular, it will become crucial for searches of ultra high-redshift LBGs to closely consider contamination from lower-redshift ($z\sim4-7$) dusty sources with significant nebular line emission that can mimic the colors of a higher-redshift Lyman break.

Despite sitting at lower redshift, new discoveries and characterizations of $z\sim5$ DSFGs will also shed new light on an otherwise mysterious population, where fewer than a few dozen systems are currently known.  Such discoveries will enable a major step forward in our understanding of massive galaxy formation in the first $\sim$1\,Gyr of the Universe's history.

\vspace{1cm}
\begin{acknowledgments}
We thank the reviewer for a constructive report that improved the clarity of our results. We also thank Jim Dunlop for helpful discussions.

V.B. and D.B. thank the Programme National de Cosmologie et Galaxies and CNES for their support. We thank Médéric Boquien and Yannick Roehlly for their help.  C.M.C. thanks the National Science Foundation for support through grants AST-1814034, and AST-2009577 and additionally the Research Corporation for Science Advancement from a 2019 Cottrell Scholar Award sponsored by IF/THEN, an initiative of Lyda Hill Philanthropies. I.A. acknowledges support from CONACyT CB-382947. We acknowledge support from STScI through award JWST-ERS-1345.

This work is based on observations made with the NASA/ESA/CSA James Webb Space Telescope. The data were obtained from the Mikulski Archive for Space Telescopes at the Space Telescope Science Institute, which is operated by the Association of Universities for Research in Astronomy, Inc., under NASA contract NAS 5-03127 for JWST. These observations are associated with program \#1345 and can be accessed in a raw format via doi:\dataset[10.17909/4abm-k128]{https://doi.org/10.17909/4abm-k128}.
This work is based on observations carried out under project number W20CK with the IRAM NOEMA Interferometer. IRAM is supported by INSU/CNRS (France), MPG (Germany) and IGN (Spain).

\end{acknowledgments}

\facilities{JWST, NOEMA, JCMT}





\appendix
\section{Assessing the reliability of high-redshift galaxy candidates via dust emission}\label{app:k-correction}

Continuum observations at submillimeter and millimeter wavelengths probe galaxies' dust thermal emission for a wide range of redshifts. Here, adopting typical dust SEDs and relationships between dust continuum emission and other physical properties, we estimate the IR luminosity, SFR, and dust mass as a function of redshift implied by a dust continuum detection similar to the one reported in this work.  Then, we compare these quantities with the expected galaxies' properties at $z>10$ to assess whether or not they lie within the realm of high-redshift galaxies.

\begin{figure*}
\centering
\includegraphics[width=\textwidth]{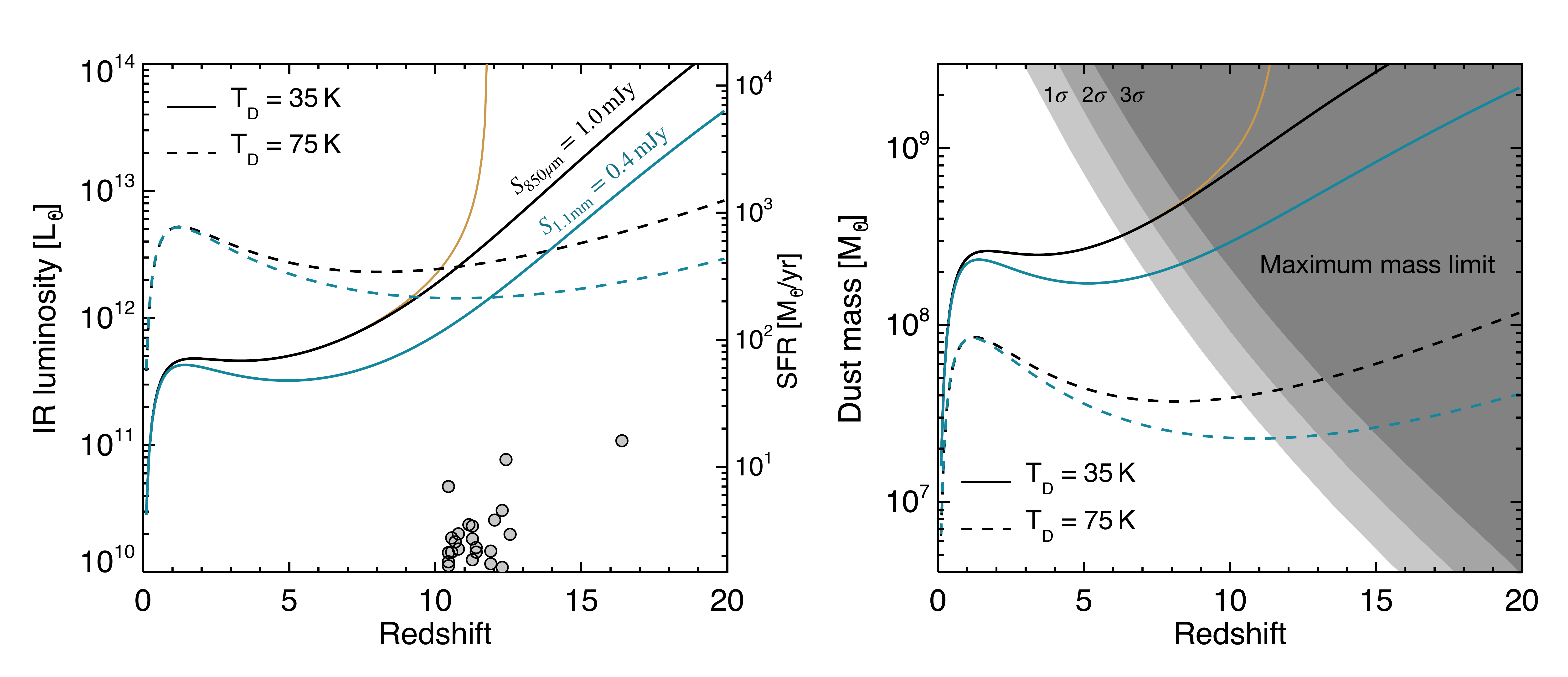}
\caption{{\it Left:} predicted IR luminosity and SFR as a function of redshift for dust continuum detections of $S_{850\,\rm\mu m}=1\,$mJy (black lines) and $S_{1,1\,\rm mm}=0.4\,$mJy (blue lines), using a modified blackbody function with a dust emissivity index of $\beta=1.8$ and dust temperature of 35 or 75\,K (solid and dashed lines, respectively). For comparison, we show the SFRs of all the $z>10$ galaxy candidates reported by \citet{Donnan2022} based on JWST/NIRCam observations in several fields. Note that the CMB effects are taken into account following \citet{daCunha2013a}. The yellow line shows the predictions ignoring the extra heating produced by CMB photons, which results in a lack of contrast between the CMB and the dust emission at any redshift greater than $T_{\rm D}=T_{\rm CMB}(z)$ (for clarity, this is only shown for the $850\,\mu\rm m$ emission and for $T_{\rm D}=35$\,K).
{\it Right:} predicted dust mass as a function of redshift for the same set of SEDs. In addition, we plot the exclusion curves at 1$\sigma$, 2$\sigma$, and 3$\sigma$ in the mass-redshift plane to illustrate the maximum dust mass of a galaxy allowed by $\Lambda$CDM cosmology within the volume covered by CEERS (see details in the main text). Based on these results, we conclude that any submillimeter/millimeter detection in the CEERS field (or any other field of similar area) with a flux density similar to that from \zavalasource\ or from the tentative emission around \donnansource\  is unlikely to come from a $z\gtrsim10$ galaxy. }
\label{fig:k-correction}.
\end{figure*}

For these calculations, we adopt a modified blackbody distribution with a dust emissivity index of $\beta=1.8$ for the dust SED (e.g. \citealt{Casey2012a}). Two different dust temperatures of $35$ and $75\,$K are explored. Then, the IR luminosity at a given redshift is estimated by, first, scaling the redshifted SED to the 850\,\um\ flux density and, second, integrating over  8–1000\,\um\ (in the rest frame). The CMB effects on the observed flux density are also taken into account following \citealt{daCunha2013a}. The inferred IR luminosity as a function of redshift for a $S_{850\,\rm\mu m}=1\,$mJy dust detection is shown in the left panel of Figure \ref{fig:k-correction}. The corresponding dust-obscured SFR estimated directly from the IR luminosity (\citealt{Kennicutt2012a}) is also indicated on the right axis. Then, we calculate the dust mass as follows. At a given redshift, we estimate the rest-frame 850\,\um\ flux density, $S_{850\,\rm\mu m, rest}$,  from the scaled SED described above (which takes into account the CMB effects) and use the following equation:
\begin{equation}
    M_{\rm d}=\frac{S_{850\rm\mu m, rest}\,D_L^2}{(1+z)\,\kappa_{\rm ref}\,B(\nu_{\rm ref},T_{\rm d})},
\end{equation}
where  $\kappa_{\rm ref}$ represents the dust mass absorption coefficient at a reference wavelength and $B(\nu_{\rm ref},T_{\rm d})$ represents the Planck function evaluated at the same frequency. We adopt $\kappa(\rm 850\,\mu m)=0.043\,\rm m^2\,kg^{-1}$ (\citealt{Li2001}) for this calculation. The implied dust mass as a function of redshift is plotted in the right panel of Figure \ref{fig:k-correction}.

As clearly seen  in Figure \ref{fig:k-correction}, a dust continuum detection of $S_{850\,\rm\mu m}\sim1\,$mJy (or $S_{1.1\,\rm mm}\sim0.4\,$mJy) at $z>10$ would imply a SFR in excess of $\sim100\,M_\odot\,\rm yr^{-1}$, rapidly reaching $\sim1,000\,M_\odot\,\rm yr^{-1}$ at $z\sim15$ (depending on the adopted temperature). This SFR is significantly higher than what is measured in any $z\gtrsim8$ object and around two orders of magnitude higher than the SFRs  inferred for JWST-selected candidates.
In the right panel of Figure \ref{fig:k-correction},  the inferred dust mass is compared with the maximum mass limit allowed by a $\Lambda$CDM universe (which depends on redshift and survey volume\footnote{The used survey area for the maximum halo mass calculation corresponds to  34.5\,arcmin$^2$, the covered area by the current CEERS/NIRCam observations.}). To estimate this limit, we use the halo mass function from  \citet{Harrison2013} scaling; first, the halo mass down by a factor of 20 (following \citealt{Marrone2018a}; see also \citealt{Casey2021a}) to approximate the corresponding galaxy ISM mass, and, second, by a factor of 100, which corresponds to the ISM-to-dust ratio  typically measured in massive galaxies (e.g. \citealt{Magdis2012a,Remy-Ruyer2014a,Scoville2016a}). As shown in the figure, within the volume probed by the CEERS observations, a submillimeter detection at $z\gtrsim6$  start to be in tension (at $1\sigma$ level) with the maximum mass limit inferred from the halo mass function (when adopting $T_{\rm d}=35\,$K). This could be slightly alleviated if the dust temperature is higher. Nevertheless, \citet{Scoville2016a} argue that, even when the luminosity-weighted dust temperature could be higher at higher redshifts (e.g. \citealt{Faisst2017,Bakx2020,Sommovigo2022}),  the mass-weighted temperature is usually cold ($\approx25-35\,K$). Furthermore, even adopting the results from this relatively high dust temperature at face value, the implied dust masses exceed the expected mass limit at $z\gtrsim12$, implying that such a system is unlikely to exist.

While these estimates represent zero-order approximations and depend strongly on the adopted assumptions (which might not be valid at very high redshifts), it is clear that dust continuum detections (on the order of $S_{850\,\rm\mu m}\sim1\,$mJy) strongly disfavor high redshifts, $z>10$, solutions for galaxies discovered in small surveys such as those conducted to date by the JWST. Submillimeter/millimeter surveys can thus be used to efficiently identify lower-redshift interlopers (i.e. dusty, star-forming galaxies) in samples of very high-redshift galaxy candidates.

\section{Extracted photometry for \donnansource}\label{appendix}
The photometry used during the SED fitting procedure on \donnansource\ is listed in Table~\ref{tab:donnanphot}.Our  fluxes are systematically brighter than those reported by \citet{Donnan2022} in all the detected bands, although the difference is small (with an average magnitude difference of $-0.09\,$mag). This could be related to the different processes used to reduce the data and the applied correction factors as discussed in \citealt{Finkelstein2022b}.

\begin{deluxetable}{ccccccccc}[h]
\tabletypesize{\small}
\tablecaption{Measured Photometry of \donnansource}
\tablewidth{\textwidth}\label{tab:donnanphot}
\tablehead{
\colhead{Instrument/Filter} & \colhead{Wavelength} & \colhead{Flux Density}\\
}
\startdata
NIRCam/F115W & 1.15\,\um & -4.1$\pm$5.9\,nJy \\
NIRCam/F150W & 1.50\,\um & 7.0$\pm$6.7\,nJy \\
NIRCam/F200W & 2.00\,\um & 22.5$\pm$4.9\,nJy \\
NIRCam/F277W & 2.77\,\um & 94.2$\pm$4.6\,nJy \\
NIRCam/F356W & 3.56\,\um & 95.8$\pm$3.7\,nJy  \\
NIRCam/F410M & 4.10\,\um & 102.4$\pm$7.3\,nJy \\
NIRCam/F444W & 4.44\,\um & 89.7$\pm$5.4\,nJy \\
{\sc Scuba-2}/850\,\um & 850\,\um & 0.65$\pm$0.26\,mJy \\
\enddata
\end{deluxetable}


\bibliography{sample631}{}

\begin{thebibliography}{}
\expandafter\ifx\csname natexlab\endcsname\relax\def\natexlab#1{#1}\fi
\providecommand{\url}[1]{\href{#1}{#1}}
\providecommand{\dodoi}[1]{doi:~\href{http://doi.org/#1}{\nolinkurl{#1}}}
\providecommand{\doeprint}[1]{\href{http://ascl.net/#1}{\nolinkurl{http://ascl.net/#1}}}
\providecommand{\doarXiv}[1]{\href{https://arxiv.org/abs/#1}{\nolinkurl{https://arxiv.org/abs/#1}}}

\bibitem[{{Adams} {et~al.}(2022){Adams}, {Conselice}, {Ferreira}, {Austin},
  {Trussler}, {Juod{\v{z}}balis}, {Wilkins}, {Caruana}, \&
  {Dayal}}]{Adams2022a}
{Adams}, N.~J., {Conselice}, C.~J., {Ferreira}, L., {et~al.} 2022, arXiv
  e-prints, arXiv:2207.11217.
\newblock \doarXiv{2207.11217}

\bibitem[{{Aravena} {et~al.}(2020){Aravena}, {Boogaard},
  {G{\'o}nzalez-L{\'o}pez}, {Decarli}, {Walter}, {Carilli}, {Smail}, {Weiss},
  {Assef}, {Bauer}, {Bouwens}, {Cortes}, {Cox}, {da Cunha}, {Daddi},
  {D{\'\i}az-Santos}, {Inami}, {Ivison}, {Novak}, {Popping}, {Riechers}, {van
  der Werf}, \& {Wagg}}]{Aravena2020a}
{Aravena}, M., {Boogaard}, L., {G{\'o}nzalez-L{\'o}pez}, J., {et~al.} 2020,
  arXiv e-prints, arXiv:2006.04284.
\newblock \doarXiv{2006.04284}

\bibitem[{{Atek} {et~al.}(2022){Atek}, {Shuntov}, {Furtak}, {Richard}, {Kneib},
  {Mahler Adi Zitrin}, \& {McCracken Clotilde Laigle St{\'e}phane
  Charlot}}]{Atek2022}
{Atek}, H., {Shuntov}, M., {Furtak}, L.~J., {et~al.} 2022, arXiv e-prints,
  arXiv:2207.12338.
\newblock \doarXiv{2207.12338}

\bibitem[{{Bagley} {et~al.}(2022){Bagley}, {Finkelstein}, {Koekemoer},
  {Ferguson}, {Arrabal Haro}, {Dickinson}, {Kartaltepe}, {Papovich},
  {P{\'e}rez-Gonz{\'a}lez}, {Pirzkal}, {Somerville}, {Willmer}, {Yang}, {Yung},
  {Fontana}, {Grazian}, {Grogin}, {Hirschmann}, {Kewley}, {Kirkpatrick},
  {Kocevski}, {Lotz}, {Medrano}, {Morales}, {Pentericci}, {Ravindranath},
  {Trump}, {Wilkins}, {Calabr{\`o}}, {Cooper}, {Costantin}, {de la Vega},
  {Hutchison}, {Lucas}, {McGrath}, {Wang}, \& {Wuyts}}]{Bagley2022}
{Bagley}, M.~B., {Finkelstein}, S.~L., {Koekemoer}, A.~M., {et~al.} 2022, arXiv
  e-prints, arXiv:2211.02495.
\newblock \doarXiv{2211.02495}

\bibitem[{{Bakx} {et~al.}(2020){Bakx}, {Tamura}, {Hashimoto}, {Inoue}, {Lee},
  {Mawatari}, {Ota}, {Umehata}, {Zackrisson}, {Hatsukade}, {Kohno}, {Matsuda},
  {Matsuo}, {Okamoto}, {Shibuya}, {Shimizu}, {Taniguchi}, \&
  {Yoshida}}]{Bakx2020}
{Bakx}, T. J.~L.~C., {Tamura}, Y., {Hashimoto}, T., {et~al.} 2020, \mnras, 493,
  4294, \dodoi{10.1093/mnras/staa509}

\bibitem[{{Barger} {et~al.}(2007){Barger}, {Cowie}, \& {Wang}}]{barger2007}
{Barger}, A.~J., {Cowie}, L.~L., \& {Wang}, W.~H. 2007, \apj, 654, 764,
  \dodoi{10.1086/509102}

\bibitem[{{Barro} {et~al.}(2011){Barro}, {P{\'e}rez-Gonz{\'a}lez}, {Gallego},
  {Ashby}, {Kajisawa}, {Miyazaki}, {Villar}, {Yamada}, \&
  {Zamorano}}]{Barro2011a}
{Barro}, G., {P{\'e}rez-Gonz{\'a}lez}, P.~G., {Gallego}, J., {et~al.} 2011,
  \apjs, 193, 13, \dodoi{10.1088/0067-0049/193/1/13}

\bibitem[{{Behroozi} {et~al.}(2019){Behroozi}, {Wechsler}, {Hearin}, \&
  {Conroy}}]{Behroozi2019a}
{Behroozi}, P., {Wechsler}, R.~H., {Hearin}, A.~P., \& {Conroy}, C. 2019,
  \mnras, 488, 3143, \dodoi{10.1093/mnras/stz1182}

\bibitem[{{Bertin} \& {Arnouts}(1996)}]{Bertin1996a}
{Bertin}, E., \& {Arnouts}, S. 1996, \aaps, 117, 393,
  \dodoi{10.1051/aas:1996164}

\bibitem[{{Boquien} {et~al.}(2019){Boquien}, {Burgarella}, {Roehlly}, {Buat},
  {Ciesla}, {Corre}, {Inoue}, \& {Salas}}]{Boquien2019}
{Boquien}, M., {Burgarella}, D., {Roehlly}, Y., {et~al.} 2019, \aap, 622, A103,
  \dodoi{10.1051/0004-6361/201834156}

\bibitem[{{Bouwens} {et~al.}(2020){Bouwens}, {Gonzalez-Lopez}, {Aravena},
  {Decarli}, {Novak}, {Stefanon}, {Walter}, {Boogaard}, {Carilli},
  {Dudzeviciute}, {Smail}, {Daddi}, {da Cunha}, {Ivison}, {Nanayakkara},
  {Cortes}, {Cox}, {Inami}, {Oesch}, {Popping}, {Riechers}, {van der Werf},
  {Weiss}, {Fudamoto}, \& {Wagg}}]{Bouwens2020a}
{Bouwens}, R., {Gonzalez-Lopez}, J., {Aravena}, M., {et~al.} 2020, arXiv
  e-prints, arXiv:2009.10727.
\newblock \doarXiv{2009.10727}

\bibitem[{{Boylan-Kolchin}(2022)}]{Boylan-Kolchin2022}
{Boylan-Kolchin}, M. 2022, arXiv e-prints, arXiv:2208.01611.
\newblock \doarXiv{2208.01611}

\bibitem[{{Brammer} {et~al.}(2008){Brammer}, {van Dokkum}, \&
  {Coppi}}]{Brammer2008a}
{Brammer}, G.~B., {van Dokkum}, P.~G., \& {Coppi}, P. 2008, \apj, 686, 1503,
  \dodoi{10.1086/591786}

\bibitem[{{Bruzual} \& {Charlot}(2003)}]{Bruzual_Charlot2003}
{Bruzual}, G., \& {Charlot}, S. 2003, \mnras, 344, 1000,
  \dodoi{10.1046/j.1365-8711.2003.06897.x}

\bibitem[{{Burgarella} {et~al.}(2005){Burgarella}, {Buat}, \&
  {Iglesias-P{\'a}ramo}}]{Burgarella2005}
{Burgarella}, D., {Buat}, V., \& {Iglesias-P{\'a}ramo}, J. 2005, \mnras, 360,
  1413, \dodoi{10.1111/j.1365-2966.2005.09131.x}

\bibitem[{{Calzetti} {et~al.}(2000){Calzetti}, {Armus}, {Bohlin}, {Kinney},
  {Koornneef}, \& {Storchi-Bergmann}}]{Calzetti2000}
{Calzetti}, D., {Armus}, L., {Bohlin}, R.~C., {et~al.} 2000, \apj, 533, 682,
  \dodoi{10.1086/308692}

\bibitem[{{Cardona-Torres} {et~al.}(2022){Cardona-Torres}, {Aretxaga},
  {Monta{\~n}a}, {Zavala}, \& {Faber}}]{Cardona-Torres2022a}
{Cardona-Torres}, L., {Aretxaga}, I., {Monta{\~n}a}, A., {Zavala}, J.~A., \&
  {Faber}, S.~M. 2022, \mnras, \dodoi{10.1093/mnras/stac2868}

\bibitem[{{Casey}(2012)}]{Casey2012a}
{Casey}, C.~M. 2012, \mnras, 425, 3094,
  \dodoi{10.1111/j.1365-2966.2012.21455.x}

\bibitem[{{Casey}(2020)}]{Casey2020a}
---. 2020, \apj, 900, 68, \dodoi{10.3847/1538-4357/aba528}

\bibitem[{{Casey} {et~al.}(2014{\natexlab{a}}){Casey}, {Narayanan}, \&
  {Cooray}}]{Casey2014a}
{Casey}, C.~M., {Narayanan}, D., \& {Cooray}, A. 2014{\natexlab{a}}, \physrep,
  541, 45, \dodoi{10.1016/j.physrep.2014.02.009}

\bibitem[{{Casey} {et~al.}(2013){Casey}, {Chen}, {Cowie}, {Barger}, {Capak},
  {Ilbert}, {Koss}, {Lee}, {Le Floc'h}, {Sanders}, \& {Williams}}]{Casey2013a}
{Casey}, C.~M., {Chen}, C.-C., {Cowie}, L.~L., {et~al.} 2013, \mnras, 436,
  1919, \dodoi{10.1093/mnras/stt1673}

\bibitem[{{Casey} {et~al.}(2014{\natexlab{b}}){Casey}, {Scoville}, {Sanders},
  {Lee}, {Cooray}, {Finkelstein}, {Capak}, {Conley}, {De Zotti}, {Farrah},
  {Fu}, {Le Floc'h}, {Ilbert}, {Ivison}, \& {Takeuchi}}]{casey2014b}
{Casey}, C.~M., {Scoville}, N.~Z., {Sanders}, D.~B., {et~al.}
  2014{\natexlab{b}}, \apj, 796, 95, \dodoi{10.1088/0004-637X/796/2/95}

\bibitem[{{Casey} {et~al.}(2018){Casey}, {Zavala}, {Spilker}, {da Cunha},
  {Hodge}, {Hung}, {Staguhn}, {Finkelstein}, \& {Drew}}]{casey2018a}
{Casey}, C.~M., {Zavala}, J.~A., {Spilker}, J., {et~al.} 2018, \apj, 862, 77,
  \dodoi{10.3847/1538-4357/aac82d}

\bibitem[{{Casey} {et~al.}(2021{\natexlab{a}}){Casey}, {Zavala}, {Manning},
  {Aravena}, {B{\'e}thermin}, {Caputi}, {Champagne}, {Clements}, {Drew},
  {Finkelstein}, {Fujimoto}, {Hayward}, {Dekel}, {Kokorev}, {Lagos}, {Long},
  {Magdis}, {Man}, {Mitsuhashi}, {Popping}, {Spilker}, {Staguhn}, {Talia},
  {Toft}, {Treister}, {Weaver}, \& {Yun}}]{Casey2021}
{Casey}, C.~M., {Zavala}, J.~A., {Manning}, S.~M., {et~al.} 2021{\natexlab{a}},
  \apj, 923, 215, \dodoi{10.3847/1538-4357/ac2eb4}

\bibitem[{{Casey} {et~al.}(2021{\natexlab{b}}){Casey}, {Zavala}, {Manning},
  {Aravena}, {B{\'e}thermin}, {Caputi}, {Champagne}, {Clements}, {Drew},
  {Finkelstein}, {Fujimoto}, {Hayward}, {Dekel}, {Kokorev}, {Lagos}, {Long},
  {Magdis}, {Man}, {Mitsuhashi}, {Popping}, {Spilker}, {Staguhn}, {Talia},
  {Toft}, {Treister}, {Weaver}, \& {Yun}}]{Casey2021a}
---. 2021{\natexlab{b}}, \apj, 923, 215, \dodoi{10.3847/1538-4357/ac2eb4}

\bibitem[{{Castellano} {et~al.}(2022){Castellano}, {Fontana}, {Treu},
  {Santini}, {Merlin}, {Leethochawalit}, {Trenti}, {Mestric}, {Vanzella},
  {Bonchi}, {Belfiori}, {Nonino}, {Paris}, {Polenta}, {Roberts-Borsani},
  {Boyett}, {Calabro}, {Glazebrook}, {Grillo}, {Mascia}, {Mason}, {Mercurio},
  {Morishita}, {Nanayakkara}, {Pentericci}, {Rosati}, {Vulcani}, {Wang}, \&
  {Yang}}]{Castellano2022}
{Castellano}, M., {Fontana}, A., {Treu}, T., {et~al.} 2022, arXiv e-prints,
  arXiv:2207.09436.
\newblock \doarXiv{2207.09436}

\bibitem[{{Chapman} {et~al.}(2003){Chapman}, {Barger}, {Cowie}, {Scott},
  {Borys}, {Capak}, {Fomalont}, {Lewis}, {Richards}, {Steffen}, {Wilson}, \&
  {Yun}}]{chapman2003}
{Chapman}, S.~C., {Barger}, A.~J., {Cowie}, L.~L., {et~al.} 2003, \apj, 585,
  57, \dodoi{10.1086/345980}

\bibitem[{{da Cunha} {et~al.}(2013){da Cunha}, {Groves}, {Walter}, {Decarli},
  {Weiss}, {Bertoldi}, {Carilli}, {Daddi}, {Elbaz}, {Ivison}, {Maiolino},
  {Riechers}, {Rix}, {Sargent}, \& {Smail}}]{daCunha2013a}
{da Cunha}, E., {Groves}, B., {Walter}, F., {et~al.} 2013, \apj, 766, 13,
  \dodoi{10.1088/0004-637X/766/1/13}

\bibitem[{{da Cunha} {et~al.}(2015){da Cunha}, {Walter}, {Smail}, {Swinbank},
  {Simpson}, {Decarli}, {Hodge}, {Weiss}, {van der Werf}, {Bertoldi},
  {Chapman}, {Cox}, {Danielson}, {Dannerbauer}, {Greve}, {Ivison}, {Karim}, \&
  {Thomson}}]{daCunha2015a}
{da Cunha}, E., {Walter}, F., {Smail}, I.~R., {et~al.} 2015, \apj, 806, 110,
  \dodoi{10.1088/0004-637X/806/1/110}

\bibitem[{{Donnan} {et~al.}(2022){Donnan}, {McLeod}, {Dunlop}, {McLure},
  {Carnall}, {Begley}, {Cullen}, {Hamadouche}, {Bowler}, {McCracken},
  {Milvang-Jensen}, {Moneti}, \& {Targett}}]{Donnan2022}
{Donnan}, C.~T., {McLeod}, D.~J., {Dunlop}, J.~S., {et~al.} 2022, arXiv
  e-prints, arXiv:2207.12356.
\newblock \doarXiv{2207.12356}

\bibitem[{{Draine} {et~al.}(2014){Draine}, {Aniano}, {Krause}, {Groves},
  {Sandstrom}, {Braun}, {Leroy}, {Klaas}, {Linz}, {Rix}, {Schinnerer},
  {Schmiedeke}, \& {Walter}}]{Draine2014}
{Draine}, B.~T., {Aniano}, G., {Krause}, O., {et~al.} 2014, \apj, 780, 172,
  \dodoi{10.1088/0004-637X/780/2/172}

\bibitem[{{Drew} \& {Casey}(2022)}]{drew2022a}
{Drew}, P.~M., \& {Casey}, C.~M. 2022, \apj, 930, 142,
  \dodoi{10.3847/1538-4357/ac6270}

\bibitem[{{Dunlop} {et~al.}(2007){Dunlop}, {Cirasuolo}, \&
  {McLure}}]{Dunlop2007}
{Dunlop}, J.~S., {Cirasuolo}, M., \& {McLure}, R.~J. 2007, \mnras, 376, 1054,
  \dodoi{10.1111/j.1365-2966.2007.11453.x}

\bibitem[{{Dwek} {et~al.}(2014){Dwek}, {Staguhn}, {Arendt}, {Kovacks}, {Su}, \&
  {Benford}}]{Dwek2014a}
{Dwek}, E., {Staguhn}, J., {Arendt}, R.~G., {et~al.} 2014, \apjl, 788, L30,
  \dodoi{10.1088/2041-8205/788/2/L30}

\bibitem[{{Dye} {et~al.}(2008){Dye}, {Eales}, {Aretxaga}, {Serjeant}, {Dunlop},
  {Babbedge}, {Chapman}, {Cirasuolo}, {Clements}, {Coppin}, {Dunne}, {Egami},
  {Farrah}, {Ivison}, {van Kampen}, {Pope}, {Priddey}, {Rieke}, {Schael},
  {Scott}, {Simpson}, {Takagi}, {Takata}, \& {Vaccari}}]{Dye2008}
{Dye}, S., {Eales}, S.~A., {Aretxaga}, I., {et~al.} 2008, \mnras, 386, 1107,
  \dodoi{10.1111/j.1365-2966.2008.13113.x}

\bibitem[{{Faisst} {et~al.}(2017){Faisst}, {Capak}, {Yan}, {Pavesi},
  {Riechers}, {Bari{\v{s}}i{\'c}}, {Cooke}, {Kartaltepe}, \&
  {Masters}}]{Faisst2017}
{Faisst}, A.~L., {Capak}, P.~L., {Yan}, L., {et~al.} 2017, \apj, 847, 21,
  \dodoi{10.3847/1538-4357/aa886c}

\bibitem[{{Finkelstein}(2016)}]{Finkelstein2016}
{Finkelstein}, S.~L. 2016, \pasa, 33, e037, \dodoi{10.1017/pasa.2016.26}

\bibitem[{{Finkelstein} {et~al.}(2017){Finkelstein}, {Dickinson}, {Ferguson},
  {Grazian}, {Grogin}, {Kartaltepe}, {Kewley}, {Kocevski}, {Koekemoer}, {Lotz},
  {Papovich}, {Pentericci}, {Perez-Gonzalez}, {Pirzkal}, {Ravindranath},
  {Somerville}, {Trump}, \& {Wilkins}}]{Finkelstein2017a}
{Finkelstein}, S.~L., {Dickinson}, M., {Ferguson}, H.~C., {et~al.} 2017, {The
  Cosmic Evolution Early Release Science (CEERS) Survey}, JWST Proposal ID
  1345. Cycle 0 Early Release Science

\bibitem[{{Finkelstein} {et~al.}(2022{\natexlab{a}}){Finkelstein}, {Bagley},
  {Arrabal Haro}, {Dickinson}, {Ferguson}, {Kartaltepe}, {Papovich},
  {Burgarella}, {Kocevski}, {Huertas-Company}, {Iyer}, {Larson},
  {P{\'e}rez-Gonz{\'a}lez}, {Rose}, {Tacchella}, {Wilkins}, {Medrano},
  {Morales}, {Somerville}, {Yung}, {Fontana}, {Giavalisco}, {Grazian},
  {Grogin}, {Kewley}, {Koekemoer}, {Kirkpatrick}, {Kurczynski}, {Lotz},
  {Pentericci}, {Pirzkal}, {Ravindranath}, {Ryan Jr.}, {Trump}, {Yang},
  {Almaini}, {Amorin}, {Annunziatella}, {Backhaus}, {Barro}, {Behroozi},
  {Bell}, {Bhatawdekar}, {Bisigello}, {Bromm}, {Buat}, {Buitrago}, {Calabro},
  {Casey}, {Castellano}, {Ch{\'a}vez Ortiz}, {Ciesla}, {Cleri}, {Cohen},
  {Cole}, {Cooke}, {Cooper}, {Cooray}, {Costantin}, {Cox}, {Croton}, {Daddi},
  {Dave}, {de la Vega}, {Dekel}, {Elbaz}, {Estrada-Carpenter}, {Faber},
  {Fern{\'a}ndez}, {Finkelstein}, {Freundlich}, {Fujimoto}, {Garcia-Argumanez},
  {Gardner}, {Gawiser}, {Gomez-Guijarro}, {Guo}, {Hamilton}, {Hathi},
  {Holwerda}, {Hirschmann}, {Hutchison}, {Jha}, {Jogee}, {Juneau}, {Jung},
  {Kassin}, {Le Bail}, {Leung}, {Lucas}, {Magnelli}, {Mantha}, {Matharu},
  {McGrath}, {McIntosh}, {Merlin}, {Mobasher}, {Newman}, {Nicholls}, {Pandya},
  {Rafelski}, {Ronayne}, {Santini}, {Seill{\'e}}, {Shah}, {Shen}, {Simons},
  {Snyder}, {Stanway}, {Straughn}, {Teplitz}, {Vanderhoof}, {Vega-Ferrero},
  {Wang}, {Weiner}, {Willmer}, {Wuyts}, \& {Zavala}}]{Finkelstein2022}
{Finkelstein}, S.~L., {Bagley}, M.~B., {Arrabal Haro}, P., {et~al.}
  2022{\natexlab{a}}, arXiv e-prints, arXiv:2207.12474.
\newblock \doarXiv{2207.12474}

\bibitem[{{Finkelstein} {et~al.}(2022{\natexlab{b}}){Finkelstein}, {Bagley},
  {Ferguson}, {Wilkins}, {Kartaltepe}, {Papovich}, {Yung}, {Arrabal Haro},
  {Behroozi}, {Dickinson}, {Kocevski}, {Koekemoer}, {Larson}, {Le Bail},
  {Morales}, {Perez-Gonzalez}, {Burgarella}, {Dave}, {Hirschmann},
  {Somerville}, {Wuyts}, {Bromm}, {Casey}, {Fontana}, {Fujimoto}, {Gardner},
  {Giavalisco}, {Grazian}, {Grogin}, {Hathi}, {Hutchison}, {Jha}, {Jogee},
  {Kewley}, {Kirkpatrick}, {Long}, {Lotz}, {Pentericci}, {Pierel}, {Pirzkal},
  {Ravindranath}, {Ryan}, {Trump}, {Yang}, {Bhatawdekar}, {Bisigello}, {Buat},
  {Calabro}, {Castellano}, {Cleri}, {Cooper}, {Croton}, {Daddi}, {Dekel},
  {Elbaz}, {Franco}, {Gawiser}, {Holwerda}, {Huertas-Company}, {Jaskot},
  {Leung}, {Lucas}, {Mobasher}, {Pandya}, {Tacchella}, {Weiner}, \&
  {Zavala}}]{Finkelstein2022b}
{Finkelstein}, S.~L., {Bagley}, M.~B., {Ferguson}, H.~C., {et~al.}
  2022{\natexlab{b}}, arXiv e-prints, arXiv:2211.05792.
\newblock \doarXiv{2211.05792}

\bibitem[{{G{\'o}mez-Guijarro} {et~al.}(2022){G{\'o}mez-Guijarro}, {Elbaz},
  {Xiao}, {Kokorev}, {Magdis}, {Magnelli}, {Daddi}, {Valentino}, {Sargent},
  {Dickinson}, {B{\'e}thermin}, {Franco}, {Pope}, {Kalita}, {Ciesla},
  {Demarco}, {Inami}, {Rujopakarn}, {Shu}, {Wang}, {Zhou}, {Alexander},
  {Bournaud}, {Chary}, {Ferguson}, {Finkelstein}, {Giavalisco}, {Iono},
  {Juneau}, {Kartaltepe}, {Lagache}, {Le Floc'h}, {Leiton}, {Leroy}, {Lin},
  {Motohara}, {Mullaney}, {Okumura}, {Pannella}, {Papovich}, \&
  {Treister}}]{Gomez-Guijarro2022}
{G{\'o}mez-Guijarro}, C., {Elbaz}, D., {Xiao}, M., {et~al.} 2022, \aap, 659,
  A196, \dodoi{10.1051/0004-6361/202142352}

\bibitem[{{Harikane} {et~al.}(2022){Harikane}, {Ouchi}, {Oguri}, {Ono},
  {Nakajima}, {Isobe}, {Umeda}, {Mawatari}, \& {Zhang}}]{Harikane2022}
{Harikane}, Y., {Ouchi}, M., {Oguri}, M., {et~al.} 2022, arXiv e-prints,
  arXiv:2208.01612.
\newblock \doarXiv{2208.01612}

\bibitem[{{Harrison} \& {Hotchkiss}(2013)}]{Harrison2013}
{Harrison}, I., \& {Hotchkiss}, S. 2013, \jcap, 2013, 022,
  \dodoi{10.1088/1475-7516/2013/07/022}

\bibitem[{{Howell} {et~al.}(2010){Howell}, {Armus}, {Mazzarella}, {Evans},
  {Surace}, {Sanders}, {Petric}, {Appleton}, {Bothun}, {Bridge}, {Chan},
  {Charmandaris}, {Frayer}, {Haan}, {Inami}, {Kim}, {Lord}, {Madore},
  {Melbourne}, {Schulz}, {U}, {Vavilkin}, {Veilleux}, \& {Xu}}]{howell2010a}
{Howell}, J.~H., {Armus}, L., {Mazzarella}, J.~M., {et~al.} 2010, \apj, 715,
  572, \dodoi{10.1088/0004-637X/715/1/572}

\bibitem[{{Inami} {et~al.}(2022){Inami}, {Algera}, {Schouws}, {Sommovigo},
  {Bouwens}, {Smit}, {Stefanon}, {Bowler}, {Endsley}, {Ferrara}, {Oesch},
  {Stark}, {Aravena}, {Barrufet}, {da Cunha}, {Dayal}, {De Looze}, {Fudamoto},
  {Gonzalez}, {Graziani}, {Hodge}, {Hygate}, {Nanayakkara}, {Pallottini},
  {Riechers}, {Schneider}, {Topping}, \& {van der Werf}}]{Inami2022}
{Inami}, H., {Algera}, H., {Schouws}, S., {et~al.} 2022, \mnras,
  \dodoi{10.1093/mnras/stac1779}

\bibitem[{{Kennicutt} \& {Evans}(2012)}]{Kennicutt2012a}
{Kennicutt}, R.~C., \& {Evans}, N.~J. 2012, \araa, 50, 531,
  \dodoi{10.1146/annurev-astro-081811-125610}

\bibitem[{{Khusanova} {et~al.}(2021){Khusanova}, {Bethermin}, {Le F{\`e}vre},
  {Capak}, {Faisst}, {Schaerer}, {Silverman}, {Cassata}, {Yan}, {Ginolfi},
  {Fudamoto}, {Loiacono}, {Amorin}, {Bardelli}, {Boquien}, {Cimatti},
  {Dessauges-Zavadsky}, {Gruppioni}, {Hathi}, {Jones}, {Koekemoer}, {Lagache},
  {Maiolino}, {Lemaux}, {Oesch}, {Pozzi}, {Riechers}, {Romano}, {Talia},
  {Toft}, {Vergani}, {Zamorani}, \& {Zucca}}]{Khusanova2021}
{Khusanova}, Y., {Bethermin}, M., {Le F{\`e}vre}, O., {et~al.} 2021, \aap, 649,
  A152, \dodoi{10.1051/0004-6361/202038944}

\bibitem[{{Koekemoer} {et~al.}(2011){Koekemoer}, {Faber}, {Ferguson}, {Grogin},
  {Kocevski}, {Koo}, {Lai}, {Lotz}, {Lucas}, {McGrath}, {Ogaz}, {Rajan},
  {Riess}, {Rodney}, {Strolger}, {Casertano}, {Castellano}, {Dahlen},
  {Dickinson}, {Dolch}, {Fontana}, {Giavalisco}, {Grazian}, {Guo}, {Hathi},
  {Huang}, {van der Wel}, {Yan}, {Acquaviva}, {Alexander}, {Almaini}, {Ashby},
  {Barden}, {Bell}, {Bournaud}, {Brown}, {Caputi}, {Cassata}, {Challis},
  {Chary}, {Cheung}, {Cirasuolo}, {Conselice}, {Roshan Cooray}, {Croton},
  {Daddi}, {Dav{\'e}}, {de Mello}, {de Ravel}, {Dekel}, {Donley}, {Dunlop},
  {Dutton}, {Elbaz}, {Fazio}, {Filippenko}, {Finkelstein}, {Frazer}, {Gardner},
  {Garnavich}, {Gawiser}, {Gruetzbauch}, {Hartley}, {H{\"a}ussler},
  {Herrington}, {Hopkins}, {Huang}, {Jha}, {Johnson}, {Kartaltepe},
  {Khostovan}, {Kirshner}, {Lani}, {Lee}, {Li}, {Madau}, {McCarthy},
  {McIntosh}, {McLure}, {McPartland}, {Mobasher}, {Moreira}, {Mortlock},
  {Moustakas}, {Mozena}, {Nandra}, {Newman}, {Nielsen}, {Niemi}, {Noeske},
  {Papovich}, {Pentericci}, {Pope}, {Primack}, {Ravindranath}, {Reddy},
  {Renzini}, {Rix}, {Robaina}, {Rosario}, {Rosati}, {Salimbeni}, {Scarlata},
  {Siana}, {Simard}, {Smidt}, {Snyder}, {Somerville}, {Spinrad}, {Straughn},
  {Telford}, {Teplitz}, {Trump}, {Vargas}, {Villforth}, {Wagner}, {Wandro},
  {Wechsler}, {Weiner}, {Wiklind}, {Wild}, {Wilson}, {Wuyts}, \&
  {Yun}}]{koekemoer2011}
{Koekemoer}, A.~M., {Faber}, S.~M., {Ferguson}, H.~C., {et~al.} 2011, \apjs,
  197, 36, \dodoi{10.1088/0067-0049/197/2/36}

\bibitem[{{Laporte} {et~al.}(2017){Laporte}, {Ellis}, {Boone}, {Bauer},
  {Qu{\'e}nard}, {Roberts-Borsani}, {Pell{\'o}}, {P{\'e}rez-Fournon}, \&
  {Streblyanska}}]{Laporte2017a}
{Laporte}, N., {Ellis}, R.~S., {Boone}, F., {et~al.} 2017, \apjl, 837, L21,
  \dodoi{10.3847/2041-8213/aa62aa}

\bibitem[{{Larson} {et~al.}(2022){Larson}, {Hutchison}, {Bagley},
  {Finkelstein}, {Yung}, {Somerville}, {Hirschmann}, {Brammer}, {Holwerda},
  {Papovich}, {Morales}, \& {Wilkins}}]{Larson2022}
{Larson}, R.~L., {Hutchison}, T.~A., {Bagley}, M., {et~al.} 2022, arXiv
  e-prints, arXiv:2211.10035.
\newblock \doarXiv{2211.10035}

\bibitem[{{Li} \& {Draine}(2001)}]{Li2001}
{Li}, A., \& {Draine}, B.~T. 2001, \apj, 554, 778, \dodoi{10.1086/323147}

\bibitem[{{Lovell} {et~al.}(2022){Lovell}, {Harrison}, {Harikane}, {Tacchella},
  \& {Wilkins}}]{Lovell2022}
{Lovell}, C.~C., {Harrison}, I., {Harikane}, Y., {Tacchella}, S., \& {Wilkins},
  S.~M. 2022, \mnras, \dodoi{10.1093/mnras/stac3224}

\bibitem[{{Lutz} {et~al.}(2011){Lutz}, {Poglitsch}, {Altieri}, {Andreani},
  {Aussel}, {Berta}, {Bongiovanni}, {Brisbin}, {Cava}, {Cepa}, {Cimatti},
  {Daddi}, {Dominguez-Sanchez}, {Elbaz}, {F{\"o}rster Schreiber}, {Genzel},
  {Grazian}, {Gruppioni}, {Harwit}, {Le Floc'h}, {Magdis}, {Magnelli},
  {Maiolino}, {Nordon}, {P{\'e}rez Garc{\'\i}a}, {Popesso}, {Pozzi},
  {Riguccini}, {Rodighiero}, {Saintonge}, {Sanchez Portal}, {Santini}, {Shao},
  {Sturm}, {Tacconi}, {Valtchanov}, {Wetzstein}, \& {Wieprecht}}]{Lutz2011a}
{Lutz}, D., {Poglitsch}, A., {Altieri}, B., {et~al.} 2011, \aap, 532, A90,
  \dodoi{10.1051/0004-6361/201117107}

\bibitem[{{Magdis} {et~al.}(2012){Magdis}, {Daddi}, {B{\'e}thermin}, {Sargent},
  {Elbaz}, {Pannella}, {Dickinson}, {Dannerbauer}, {da Cunha}, {Walter},
  {Rigopoulou}, {Charmandaris}, {Hwang}, \& {Kartaltepe}}]{Magdis2012a}
{Magdis}, G.~E., {Daddi}, E., {B{\'e}thermin}, M., {et~al.} 2012, \apj, 760, 6,
  \dodoi{10.1088/0004-637X/760/1/6}

\bibitem[{{Magnelli} {et~al.}(2009){Magnelli}, {Elbaz}, {Chary}, {Dickinson},
  {Le Borgne}, {Frayer}, \& {Willmer}}]{Magnelli2009}
{Magnelli}, B., {Elbaz}, D., {Chary}, R.~R., {et~al.} 2009, \aap, 496, 57,
  \dodoi{10.1051/0004-6361:200811443}

\bibitem[{{Marrone} {et~al.}(2018){Marrone}, {Spilker}, {Hayward}, {Vieira},
  {Aravena}, {Ashby}, {Bayliss}, {B{\'e}thermin}, {Brodwin}, {Bothwell},
  {Carlstrom}, {Chapman}, {Chen}, {Crawford}, {Cunningham}, {De Breuck},
  {Fassnacht}, {Gonzalez}, {Greve}, {Hezaveh}, {Lacaille}, {Litke}, {Lower},
  {Ma}, {Malkan}, {Miller}, {Morningstar}, {Murphy}, {Narayanan}, {Phadke},
  {Rotermund}, {Sreevani}, {Stalder}, {Stark}, {Strand et}, {Tang}, \&
  {Wei{\ss}}}]{Marrone2018a}
{Marrone}, D.~P., {Spilker}, J.~S., {Hayward}, C.~C., {et~al.} 2018, \nat, 553,
  51, \dodoi{10.1038/nature24629}

\bibitem[{{Naidu} {et~al.}(2022){Naidu}, {Oesch}, {van Dokkum}, {Nelson},
  {Suess}, {Whitaker}, {Allen}, {Bezanson}, {Bouwens}, {Brammer}, {Conroy},
  {Illingworth}, {Labbe}, {Leja}, {Leonova}, {Matthee}, {Price}, {Setton},
  {Strait}, {Stefanon}, {Tacchella}, {Toft}, {Weaver}, \& {Weibel}}]{Naidu2022}
{Naidu}, R.~P., {Oesch}, P.~A., {van Dokkum}, P., {et~al.} 2022, arXiv
  e-prints, arXiv:2207.09434.
\newblock \doarXiv{2207.09434}

\bibitem[{{Noll} {et~al.}(2009){Noll}, {Burgarella}, {Giovannoli}, {Buat},
  {Marcillac}, \& {Mu{\~n}oz-Mateos}}]{Noll2009}
{Noll}, S., {Burgarella}, D., {Giovannoli}, E., {et~al.} 2009, \aap, 507, 1793,
  \dodoi{10.1051/0004-6361/200912497}

\bibitem[{{Oliver} {et~al.}(2012){Oliver}, {Bock}, {Altieri}, {Amblard},
  {Arumugam}, {Aussel}, {Babbedge}, {Beelen}, {B{\'e}thermin}, {Blain},
  {Boselli}, {Bridge}, {Brisbin}, {Buat}, {Burgarella},
  {Castro-Rodr{\'\i}guez}, {Cava}, {Chanial}, {Cirasuolo}, {Clements},
  {Conley}, {Conversi}, {Cooray}, {Dowell}, {Dubois}, {Dwek}, {Dye}, {Eales},
  {Elbaz}, {Farrah}, {Feltre}, {Ferrero}, {Fiolet}, {Fox}, {Franceschini},
  {Gear}, {Giovannoli}, {Glenn}, {Gong}, {Gonz{\'a}lez Solares}, {Griffin},
  {Halpern}, {Harwit}, {Hatziminaoglou}, {Heinis}, {Hurley}, {Hwang}, {Hyde},
  {Ibar}, {Ilbert}, {Isaak}, {Ivison}, {Lagache}, {Le Floc'h}, {Levenson},
  {Faro}, {Lu}, {Madden}, {Maffei}, {Magdis}, {Mainetti}, {Marchetti},
  {Marsden}, {Marshall}, {Mortier}, {Nguyen}, {O'Halloran}, {Omont}, {Page},
  {Panuzzo}, {Papageorgiou}, {Patel}, {Pearson}, {P{\'e}rez-Fournon}, {Pohlen},
  {Rawlings}, {Raymond}, {Rigopoulou}, {Riguccini}, {Rizzo}, {Rodighiero},
  {Roseboom}, {Rowan-Robinson}, {S{\'a}nchez Portal}, {Schulz}, {Scott},
  {Seymour}, {Shupe}, {Smith}, {Stevens}, {Symeonidis}, {Trichas}, {Tugwell},
  {Vaccari}, {Valtchanov}, {Vieira}, {Viero}, {Vigroux}, {Wang}, {Ward},
  {Wardlow}, {Wright}, {Xu}, \& {Zemcov}}]{oliver2012a}
{Oliver}, S.~J., {Bock}, J., {Altieri}, B., {et~al.} 2012, \mnras, 424, 1614,
  \dodoi{10.1111/j.1365-2966.2012.20912.x}

\bibitem[{{Pei}(1992)}]{Pei1992a}
{Pei}, Y.~C. 1992, \apj, 395, 130, \dodoi{10.1086/171637}

\bibitem[{{P{\'e}rez-Gonz{\'a}lez} {et~al.}(2022){P{\'e}rez-Gonz{\'a}lez},
  {Barro}, {Annunziatella}, {Costantin}, {Garc{\'\i}a-Argum{\'a}nez},
  {McGrath}, {M{\'e}rida}, {Zavala}, {Arrabal Haro}, {Bagley}, {Backhaus},
  {Behroozi}, {Bell}, {Buat}, {Calabr{\`o}}, {Casey}, {Cleri}, {Coogan},
  {Cooper}, {Cooray}, {Dekel}, {Dickinson}, {Elbaz}, {Ferguson}, {Finkelstein},
  {Fontana}, {Franco}, {Gardner}, {Giavalisco}, {G{\'o}mez-Guijarro},
  {Grazian}, {Grogin}, {Guo}, {Jogee}, {Kartaltepe}, {Kewley}, {Kirkpatrick},
  {Kocevski}, {Koekemoer}, {Long}, {Lotz}, {Lucas}, {Papovich}, {Pirzkal},
  {Ravindranath}, {Somerville}, {Tacchella}, {Trump}, {Wang}, {Wilkins},
  {Wuyts}, {Yang}, \& {Yung}}]{Perez-Gonzalez2022}
{P{\'e}rez-Gonz{\'a}lez}, P.~G., {Barro}, G., {Annunziatella}, M., {et~al.}
  2022, arXiv e-prints, arXiv:2211.00045.
\newblock \doarXiv{2211.00045}

\bibitem[{{Planck Collaboration} {et~al.}(2016){Planck Collaboration}, {Ade},
  {Aghanim}, {Arnaud}, {Ashdown}, {Aumont}, {Baccigalupi}, {Banday},
  {Barreiro}, {Bartlett}, {Bartolo}, {Battaner}, {Battye}, {Benabed},
  {Beno{\^\i}t}, {Benoit-L{\'e}vy}, {Bernard}, {Bersanelli}, {Bielewicz},
  {Bock}, {Bonaldi}, {Bonavera}, {Bond}, {Borrill}, {Bouchet}, {Boulanger},
  {Bucher}, {Burigana}, {Butler}, {Calabrese}, {Cardoso}, {Catalano},
  {Challinor}, {Chamballu}, {Chary}, {Chiang}, {Chluba}, {Christensen},
  {Church}, {Clements}, {Colombi}, {Colombo}, {Combet}, {Coulais}, {Crill},
  {Curto}, {Cuttaia}, {Danese}, {Davies}, {Davis}, {de Bernardis}, {de Rosa},
  {de Zotti}, {Delabrouille}, {D{\'e}sert}, {Di Valentino}, {Dickinson},
  {Diego}, {Dolag}, {Dole}, {Donzelli}, {Dor{\'e}}, {Douspis}, {Ducout},
  {Dunkley}, {Dupac}, {Efstathiou}, {Elsner}, {En{\ss}lin}, {Eriksen},
  {Farhang}, {Fergusson}, {Finelli}, {Forni}, {Frailis}, {Fraisse},
  {Franceschi}, {Frejsel}, {Galeotta}, {Galli}, {Ganga}, {Gauthier}, {Gerbino},
  {Ghosh}, {Giard}, {Giraud-H{\'e}raud}, {Giusarma}, {Gjerl{\o}w},
  {Gonz{\'a}lez-Nuevo}, {G{\'o}rski}, {Gratton}, {Gregorio}, {Gruppuso},
  {Gudmundsson}, {Hamann}, {Hansen}, {Hanson}, {Harrison}, {Helou},
  {Henrot-Versill{\'e}}, {Hern{\'a}ndez-Monteagudo}, {Herranz}, {Hildebrand t},
  {Hivon}, {Hobson}, {Holmes}, {Hornstrup}, {Hovest}, {Huang}, {Huffenberger},
  {Hurier}, {Jaffe}, {Jaffe}, {Jones}, {Juvela}, {Keih{\"a}nen}, {Keskitalo},
  {Kisner}, {Kneissl}, {Knoche}, {Knox}, {Kunz}, {Kurki-Suonio}, {Lagache},
  {L{\"a}hteenm{\"a}ki}, {Lamarre}, {Lasenby}, {Lattanzi}, {Lawrence}, {Leahy},
  {Leonardi}, {Lesgourgues}, {Levrier}, {Lewis}, {Liguori}, {Lilje},
  {Linden-V{\o}rnle}, {L{\'o}pez-Caniego}, {Lubin}, {Mac{\'\i}as-P{\'e}rez},
  {Maggio}, {Maino}, {Mandolesi}, {Mangilli}, {Marchini}, {Maris}, {Martin},
  {Martinelli}, {Mart{\'\i}nez-Gonz{\'a}lez}, {Masi}, {Matarrese}, {McGehee},
  {Meinhold}, {Melchiorri}, {Melin}, {Mendes}, {Mennella}, {Migliaccio},
  {Millea}, {Mitra}, {Miville-Desch{\^e}nes}, {Moneti}, {Montier}, {Morgante},
  {Mortlock}, {Moss}, {Munshi}, {Murphy}, {Naselsky}, {Nati}, {Natoli},
  {Netterfield}, {N{\o}rgaard-Nielsen}, {Noviello}, {Novikov}, {Novikov},
  {Oxborrow}, {Paci}, {Pagano}, {Pajot}, {Paladini}, {Paoletti}, {Partridge},
  {Pasian}, {Patanchon}, {Pearson}, {Perdereau}, {Perotto}, {Perrotta},
  {Pettorino}, {Piacentini}, {Piat}, {Pierpaoli}, {Pietrobon}, {Plaszczynski},
  {Pointecouteau}, {Polenta}, {Popa}, {Pratt}, {Pr{\'e}zeau}, {Prunet},
  {Puget}, {Rachen}, {Reach}, {Rebolo}, {Reinecke}, {Remazeilles}, {Renault},
  {Renzi}, {Ristorcelli}, {Rocha}, {Rosset}, {Rossetti}, {Roudier},
  {Rouill{\'e} d'Orfeuil}, {Rowan-Robinson}, {Rubi{\~n}o-Mart{\'\i}n},
  {Rusholme}, {Said}, {Salvatelli}, {Salvati}, {Sandri}, {Santos},
  {Savelainen}, {Savini}, {Scott}, {Seiffert}, {Serra}, {Shellard}, {Spencer},
  {Spinelli}, {Stolyarov}, {Stompor}, {Sudiwala}, {Sunyaev}, {Sutton},
  {Suur-Uski}, {Sygnet}, {Tauber}, {Terenzi}, {Toffolatti}, {Tomasi},
  {Tristram}, {Trombetti}, {Tucci}, {Tuovinen}, {T{\"u}rler}, {Umana},
  {Valenziano}, {Valiviita}, {Van Tent}, {Vielva}, {Villa}, {Wade}, {Wandelt},
  {Wehus}, {White}, {White}, {Wilkinson}, {Yvon}, {Zacchei}, \&
  {Zonca}}]{Planck2016a}
{Planck Collaboration}, {Ade}, P.~A.~R., {Aghanim}, N., {et~al.} 2016, \aap,
  594, A13, \dodoi{10.1051/0004-6361/201525830}

\bibitem[{{Pope} {et~al.}(2006){Pope}, {Scott}, {Dickinson}, {Chary},
  {Morrison}, {Borys}, {Sajina}, {Alexander}, {Daddi}, {Frayer}, {MacDonald},
  \& {Stern}}]{Pope2006}
{Pope}, A., {Scott}, D., {Dickinson}, M., {et~al.} 2006, \mnras, 370, 1185,
  \dodoi{10.1111/j.1365-2966.2006.10575.x}

\bibitem[{{R{\'e}my-Ruyer} {et~al.}(2014){R{\'e}my-Ruyer}, {Madden},
  {Galliano}, {Galametz}, {Takeuchi}, {Asano}, {Zhukovska}, {Lebouteiller},
  {Cormier}, {Jones}, {Bocchio}, {Baes}, {Bendo}, {Boquien}, {Boselli},
  {DeLooze}, {Doublier-Pritchard}, {Hughes}, {Karczewski}, \&
  {Spinoglio}}]{Remy-Ruyer2014a}
{R{\'e}my-Ruyer}, A., {Madden}, S.~C., {Galliano}, F., {et~al.} 2014, \aap,
  563, A31, \dodoi{10.1051/0004-6361/201322803}

\bibitem[{{Rigby} {et~al.}(2022){Rigby}, {Perrin}, {McElwain}, {Kimble},
  {Friedman}, {Lallo}, {Doyon}, {Feinberg}, {Ferruit}, {Glasse}, \&
  et~al.}]{Rigby2022a}
{Rigby}, J., {Perrin}, M., {McElwain}, M., {et~al.} 2022, arXiv e-prints,
  arXiv:2207.05632.
\newblock \doarXiv{2207.05632}

\bibitem[{{Robertson}(2021)}]{Robertson2021}
{Robertson}, B.~E. 2021, arXiv e-prints, arXiv:2110.13160.
\newblock \doarXiv{2110.13160}

\bibitem[{{Scoville} {et~al.}(2016){Scoville}, {Sheth}, {Aussel}, {Vanden
  Bout}, {Capak}, {Bongiorno}, {Casey}, {Murchikova}, {Koda},
  {{\'A}lvarez-M{\'a}rquez}, {Lee}, {Laigle}, {McCracken}, {Ilbert}, {Pope},
  {Sanders}, {Chu}, {Toft}, {Ivison}, \& {Manohar}}]{Scoville2016a}
{Scoville}, N., {Sheth}, K., {Aussel}, H., {et~al.} 2016, \apj, 820, 83,
  \dodoi{10.3847/0004-637X/820/2/83}

\bibitem[{{Simpson} {et~al.}(2017){Simpson}, {Smail}, {Swinbank}, {Ivison},
  {Dunlop}, {Geach}, {Almaini}, {Arumugam}, {Bremer}, {Chen}, {Conselice},
  {Coppin}, {Farrah}, {Ibar}, {Hartley}, {Ma}, {Micha{\l}owski}, {Scott},
  {Spaans}, {Thomson}, \& {van der Werf}}]{Simpson2017}
{Simpson}, J.~M., {Smail}, I., {Swinbank}, A.~M., {et~al.} 2017, \apj, 839, 58,
  \dodoi{10.3847/1538-4357/aa65d0}

\bibitem[{{Sommovigo} {et~al.}(2022){Sommovigo}, {Ferrara}, {Pallottini},
  {Dayal}, {Bouwens}, {Smit}, {da Cunha}, {De Looze}, {Bowler}, {Hodge},
  {Inami}, {Oesch}, {Endsley}, {Gonzalez}, {Schouws}, {Stark}, {Stefanon},
  {Aravena}, {Graziani}, {Riechers}, {Schneider}, {van der Werf}, {Algera},
  {Barrufet}, {Fudamoto}, {Hygate}, {Labb{\'e}}, {Li}, {Nanayakkara}, \&
  {Topping}}]{Sommovigo2022}
{Sommovigo}, L., {Ferrara}, A., {Pallottini}, A., {et~al.} 2022, \mnras, 513,
  3122, \dodoi{10.1093/mnras/stac302}

\bibitem[{{Stark}(2016)}]{Stark2016}
{Stark}, D.~P. 2016, \araa, 54, 761,
  \dodoi{10.1146/annurev-astro-081915-023417}

\bibitem[{{Strandet} {et~al.}(2017){Strandet}, {Weiss}, {De Breuck}, {Marrone},
  {Vieira}, {Aravena}, {Ashby}, {B{\'e}thermin}, {Bothwell}, {Bradford},
  {Carlstrom}, {Chapman}, {Cunningham}, {Chen}, {Fassnacht}, {Gonzalez},
  {Greve}, {Gullberg}, {Hayward}, {Hezaveh}, {Litke}, {Ma}, {Malkan}, {Menten},
  {Miller}, {Murphy}, {Narayanan}, {Phadke}, {Rotermund}, {Spilker}, \&
  {Sreevani}}]{Strandet2017a}
{Strandet}, M.~L., {Weiss}, A., {De Breuck}, C., {et~al.} 2017, \apjl, 842,
  L15, \dodoi{10.3847/2041-8213/aa74b0}

\bibitem[{{Tamura} {et~al.}(2019){Tamura}, {Mawatari}, {Hashimoto}, {Inoue},
  {Zackrisson}, {Christensen}, {Binggeli}, {Matsuda}, {Matsuo}, {Takeuchi},
  {Asano}, {Sunaga}, {Shimizu}, {Okamoto}, {Yoshida}, {Lee}, {Shibuya},
  {Taniguchi}, {Umehata}, {Hatsukade}, {Kohno}, \& {Ota}}]{Tamura2019a}
{Tamura}, Y., {Mawatari}, K., {Hashimoto}, T., {et~al.} 2019, \apj, 874, 27,
  \dodoi{10.3847/1538-4357/ab0374}

\bibitem[{{Wang} {et~al.}(2019){Wang}, {Schreiber}, {Elbaz}, {Yoshimura},
  {Kohno}, {Shu}, {Yamaguchi}, {Pannella}, {Franco}, {Huang}, {Lim}, \&
  {Wang}}]{Wang2019a}
{Wang}, T., {Schreiber}, C., {Elbaz}, D., {et~al.} 2019, \nat, 572, 211,
  \dodoi{10.1038/s41586-019-1452-4}

\bibitem[{{Yan} {et~al.}(2022){Yan}, {Ma}, {Ling}, {Cheng}, {Huang}, \&
  {Zitrin}}]{Yan2022}
{Yan}, H., {Ma}, Z., {Ling}, C., {et~al.} 2022, arXiv e-prints,
  arXiv:2207.11558.
\newblock \doarXiv{2207.11558}

\bibitem[{{Zavala} {et~al.}(2017){Zavala}, {Aretxaga}, {Geach}, {Hughes},
  {Birkinshaw}, {Chapin}, {Chapman}, {Chen}, {Clements}, {Dunlop}, {Farrah},
  {Ivison}, {Jenness}, {Micha{\l}owski}, {Robson}, {Scott}, {Simpson},
  {Spaans}, \& {van der Werf}}]{Zavala2017a}
{Zavala}, J.~A., {Aretxaga}, I., {Geach}, J.~E., {et~al.} 2017, \mnras, 464,
  3369, \dodoi{10.1093/mnras/stw2630}

\bibitem[{{Zavala} {et~al.}(2018{\natexlab{a}}){Zavala}, {Monta{\~n}a},
  {Hughes}, {Yun}, {Ivison}, {Valiante}, {Wilner}, {Spilker}, {Aretxaga},
  {Eales}, {Avila-Reese}, {Ch{\'a}vez}, {Cooray}, {Dannerbauer}, {Dunlop},
  {Dunne}, {G{\'o}mez-Ruiz}, {Micha{\l}owski}, {Narayanan}, {Nayyeri}, {Oteo},
  {Rosa Gonz{\'a}lez}, {S{\'a}nchez-Arg{\"u}elles}, {Schloerb}, {Serjeant},
  {Smith}, {Terlevich}, {Vega}, {Villalba}, {van der Werf}, {Wilson}, \&
  {Zeballos}}]{Zavala2018a}
{Zavala}, J.~A., {Monta{\~n}a}, A., {Hughes}, D.~H., {et~al.}
  2018{\natexlab{a}}, Nature Astronomy, 2, 56,
  \dodoi{10.1038/s41550-017-0297-8}

\bibitem[{{Zavala} {et~al.}(2018{\natexlab{b}}){Zavala}, {Aretxaga}, {Dunlop},
  {Micha{\l}owski}, {Hughes}, {Bourne}, {Chapin}, {Cowley}, {Farrah}, {Lacey},
  {Targett}, \& {van der Werf}}]{Zavala2018b}
{Zavala}, J.~A., {Aretxaga}, I., {Dunlop}, J.~S., {et~al.} 2018{\natexlab{b}},
  \mnras, 475, 5585, \dodoi{10.1093/mnras/sty217}

\end{thebibliography}
\bibliographystyle{aasjournal}



\allauthors

\end{document}